\documentclass[12pt,preprint]{aastex}

\newcommand{\bq}{\begin{equation}}
\newcommand{\eq}{\end{equation}}

\newcommand{\ffas}{\hbox{$\,.\!\!^{\prime\prime}$}}

\shorttitle{Submillimeter galaxies in GOODS--S}

\begin{document}

\title{Properties of submillimeter galaxies in the CANDELS GOODS--South Field}
\author{
Tommy Wiklind\altaffilmark{1},
Christopher J. Conselice\altaffilmark{2},
Tomas Dahlen\altaffilmark{3},
Mark E. Dickinson\altaffilmark{4},
Henry C. Ferguson\altaffilmark{3},
Norman A. Grogin\altaffilmark{3},
Yicheng Guo\altaffilmark{5},
Anton M. Koekemoer\altaffilmark{3},
Bahram Mobasher\altaffilmark{6},
Alice Mortlock\altaffilmark{2},
Adriano Fontana\altaffilmark{7},
Romeel Dav{\'e}\altaffilmark{8,9,10},
Haojing Yan\altaffilmark{11},
Viviana Acquaviva\altaffilmark{12},
Matthew L.N. Ashby\altaffilmark{13},
Guillermo Barro\altaffilmark{5},
Karina I. Caputi\altaffilmark{14},
Marco Castellano\altaffilmark{7},
Avishai Dekel\altaffilmark{15},
Jennifer L. Donley\altaffilmark{16},
Giovanni G. Fazio\altaffilmark{13},
Mauro Giavalisco\altaffilmark{17},
Andrea Grazian\altaffilmark{7},
Nimish P. Hathi\altaffilmark{18},
Peter Kurczynski\altaffilmark{19},
Yu Lu\altaffilmark{20},
Elizabeth J. McGrath\altaffilmark{21},
Duilia F. de Mello\altaffilmark{22},
Michael Peth\altaffilmark{23},
Mohammad Safarzadeh\altaffilmark{23},
Mauro Stefanon\altaffilmark{11},
Thomas Targett\altaffilmark{24,25}
}

\altaffiltext{1}{European Southern Observatory/Joint ALMA Observatory, 3107 Alonso de Cordova, Santiago, Chile; twiklind@alma.cl}
\altaffiltext{2}{Department of Physics and Astronomy, University of Nottingham, Nottingham, UK}
\altaffiltext{3}{Space Telescope Science Institute, 3700 San Martin Drive, Baltimore, MD 21218, USA}
\altaffiltext{4}{National Optical Observatory, 950 North Cherry Avenue, Tucson, AZ 85719, USA}
\altaffiltext{5}{UCO/Lick Observatory, Department of Astronomy and Astrophysics, University of California, Santa Cruz, CA 95064, USA}
\altaffiltext{6}{Department of Physics and Astronomy, University of California, Riverside, CA 95064, USA}
\altaffiltext{7}{INAF - Osservatorio Astronomico di Roma, Via Frascati 33, I00040, Monteporzio, Italy}
\altaffiltext{8}{University of the Western Cape, Bellville, Cape Town 7535, South Africa}
\altaffiltext{9}{South African Astronomical Observatories, Observatory, Cape Town 7925, South Africa}
\altaffiltext{10}{African Institute for Mathematical Sciences, Muizenberg, Cape Town 7945, South Africa}
\altaffiltext{11}{Department of Physics and Astronomy, University of Missouri, Columbia, MO 65211}
\altaffiltext{12}{Physics Department, CUNY New York City College of Technology, 300 Jay Street, Brooklyn, NY 11201, USA}
\altaffiltext{13}{Harvard-Smithsonian Center for Astrophysics, 60 Garden Street, Cambridge, MA 02138, USA}
\altaffiltext{14}{Kapteyn Astronomical Institute, University of Groningen, Groningen, The Netherlands}
\altaffiltext{15}{Center for Astrophysics and Planetary Science, Racah Institute of Physics, The Hebrew University, Jerusalem 91904, Israel}
\altaffiltext{16}{Los Alamos National Laboratory, P.O. Bx 1663, MST 087, Los Alamos, NM 87545, USA}
\altaffiltext{17}{Department of Physics and Astronomy, University of Massachusetts, 710 North Pleasant Street, Amherst, MA 01003, USA}
\altaffiltext{18} {Aix Marseille Universit\'{e}, CNRS, Laboratoire d'Astrophysique de Marseille, Marseille, France}
\altaffiltext{19}{Department of Physics and Astronomy, Rutgers, The State University of New Jersey, 136 Freulinghuysen Road, Piscataway, NJ 08852, USA}
\altaffiltext{20}{Kavli Institute for Particle Astrophysics \& Cosmology, 452 Lomita Mall, Stanford, CA 94305, USA}
\altaffiltext{21}{Department of Physics and Astronomy, Colby College, Waterville, ME 04901, USA}
\altaffiltext{22}{Department of Physics and Astronomy, Catholic University of America, 200 Hannan Hall, Washington, DC 20064, USA}
\altaffiltext{23}{Department of Physics and Astronomy, Johns Hopkins University, 3400 North Charles Street, Baltimore, MD 21218, USA}
\altaffiltext{24}{Institute for Astronomy, University of Edinburgh, Royal Observatory, Edinburgh, UK}
\altaffiltext{25}{Department of Physics and Astronomy, Sonoma State University, Rohnert Park, CA, 94928}

\begin{abstract}
We derive physical properties of 10 submillimeter galaxies located in the CANDELS coverage of the GOODS-S field.
The galaxies were first identified as submillimeter sources with the LABOCA bolometer and subsequently targeted for
870$\mu$m continuum observation with ALMA. The high angular resolution of the ALMA imaging allows secure counterparts
to be identified in the CANDELS multiband dataset. The CANDELS data provide deep photometric data from UV through
near-infrared wavelengths. Using synthetic spectral energy distributions, we derive photometric redshifts, stellar masses,
extinction, ages, and the star formation history. The redshift range is z$=$1.65-4.76, with two of the galaxies located at z$>$4.
Two SMG counterparts have stellar masses 2-3 orders of magnitude lower than the rest. The remaining SMG counterparts
have stellar masses around $1\times10^{11}$\,M$_{\odot}$. The stellar population in the SMGs is typically older than the
expected duration of the submillimeter phase, suggesting that the star formation history of submillimeter galaxies is more
complex than a single burst. Non-parametric morphology indices suggest that the SMG counterparts are among the most
asymmetric systems compared with galaxies of the same stellar mass and redshift. The HST images shows that 3 of the
SMGs are associated with on-going mergers. The remaining counterparts are isolated. Estimating the dust and molecular
gas mass from the submm fluxes, and comparing with our stellar masses shows that the gas mass fraction of SMGs is
$\sim$28\% and that the final stellar mass is likely to be $\sim(1-2)\times10^{11}$\,M$_{\odot}$.
\end{abstract}

\keywords{cosmology: observations --- galaxies: formation --- galaxies: high redshift --
galaxies: photometry --- galaxies: evolution}

\section{Introduction}\label{intro}

Submillimeter galaxies (SMGs) are the most powerful starbursts known in the universe. With
star formation rates, derived from their far-infrared luminosity, often exceeding several $10^{3}$\,M$_{\odot}$\,yr$^{-1}$, 
SMGs are capable of building up a stellar mass of $\sim10^{11}$\,M$_{\odot}$ in less than 100 Myr. 
This formation process of SMGs resembles the monolithic collapse scenario (Eggen et al. 1962), but might be
driven by hierarchical processes that merges two gas rich galaxies. The role of accretion and minor mergers is
currently unknown, but could play a role in the slightly less luminous SMGs. It has been proposed that the end
result of the SMG process are the giant elliptical galaxies seen in the local universe (Blain et al. 2002).

Despite a substantial observational effort over the last 15 years, there are still unanswered questions concerning
SMGs: the redshift distribution, the stellar mass content, what are the progenitors and descendants,  and finally,
whether the SMGs play a significant role in the overall formation and evolution of galaxies or do they just represent
the most extreme manifestation of a general star forming galaxy population?

The main hurdle in learning more about the SMG population has from the start been the difficulty in identifying
optical and near-infrared counterparts. SMGs are initially identified using single dish bolometer arrays, with an angular
resolution $\ga10''$. By pushing interferometric radio continuum observations to levels where massive star formation
is the main contributor to the continuum emission, and with sub-arcscond angular resolution, it was possible to identify
the most likely counterparts of some SMGs and thereby obtain optical spectroscopic observations (e.g. Chapman et al.
2003; Aretxaga et al. 2007; Smolcic et al. 2012a; Hodge et al. 2013). This method, however, can at best only
account for a limited fraction of all SMGs. A significant number of SMGs remain undetected even in the deepest radio continuum
observations, possibly due to being at high redshift where the radio continuum falls below the detection limit,
and some do not reveal optical emission and no spectroscopy can be done. An illustrative example of an elusive SMG 
 is the brightest submillimeter source in the Hubble Deep Field North (Hughes et al. 1998). Despite a large observational
 effort, no optical/near-infrared counterpart has been identified. The redshift was only recently determined using molecular
 emission lines and found to be $z\approx5.2$ (Walter et al. 2012). The counterpart remains undetected at other wavelengths.
In addition to using radio continuum to identify optical/NIR counterparts, Spitzer mid-IR data has also been used (e.g.
 Pope et al. 2006; Biggs et al. 2011; Targett et al. 2013; Alberts et al. 2013).

To avoid the potential selection bias inherent in using the radio continuum to identify SMG counterparts, and to achieve
a more direct association between the SMG and potential counterparts, it is better
to directly target the millimeter and submillimeter emission using interferometric techniques with sub-arcsecond
resolution. This method has successfully been used with the SMA, CARMA, Plateau de Bure and now also ALMA
 interferometers (e.g. Younger et al. 2008; Capak et al. 2011; Smolcic et al. 2012a, 2012b; Bothwell et al. 2013;
 Vieira et al. 2013; Hodge et al. 2013).
These bias-free studies have found several SMGs at redshifts $z>4$, challenging both the previously derived redshift
distribution of SMGs and the modeling of the formation of massive galaxies in the early universe (e.g. Baugh et al. 2005).
These results suggests that a substantial part of the SMG population, $\sim$30\%, is located at
$z \ga 4$ (Capak et al. 2008; Coppin et al. 2009; Daddi et al. 2009a, 2009b; Cox et al. 2011; Smolcic et al. 2011, 2012a;
Wagg et al. 2012; Swinbank et al. 2012), with at least five SMGs found at $z > 5$
(Capak et al. 2011; Combes et al. 2012; Walter et al. 2012; Vieira et al. 2013). A recent detection of a z$=$6.3 SMG was
reported by Riechers et al. (2013), pushing the redshift limit of the SMG phenomenon to within $\sim$1 Gyr from the
Big Bang. Observations of CO emission lines in lensed SMGs detected with the South Pole Telescope (Vieira et al. 2013)
suggests that the radio continuum identification of SMGs may have missed a significant fraction of $z\ga4$ SMGs
and there are indications that sources selected at longer wavelengths may have redshift distributions that peak at higher
redshifts than 870$\mu$m selected samples (e.g. Smolcic et al. 2012; Yun et al. 2012; Vieira et al. 2013).

Regardless of their redshift, SMGs tend to have a remarkably uniform molecular gas mass. Summarizing various CO
detections from the literature, and adopting a uniform CO-H$_{2}$ conversion factor of 0.8\,M$_{\odot}$\,(K\,km\,s$^{-1})^{-1}$,
the typical molecular gas mass of SMGs is $\sim3 \times 10^{10}$\,M$_{\odot}$ with only a small dispersion.
(e.g. Greve et al. 2005; Tacconi et al. 2008; Coppin et al. 2009; Bothwell et al. 2013 and references therein)

With the new ALMA interferometer it is possible to detect sub-millimeter continuum emission at a flux level of $\sim1$\,mJy
with sub-arcsecond resolution in a matter of minutes. This will greatly simplify the identification of SMG counterparts
and allow an unbiased study of the SMG population.
The first results for SMGs observed with ALMA during the Early Science phase have recently been presented by
Hodge et al. (2013) and Vieira et al. (2013). The ALMA observations of Hodge et al. (2013) targeted the 126
LESS\footnote{LESS = LABOCA ECDF Submillimeter Survey; ECDF = Extended Chandra Deep Field. The sources
observed with ALMA has been referred to as ALMA LESS, or ALESS. In the text we will for simpicity refer to the SMGs
as LESS `NN' and use the designation of Hodge et al. (2013) for multiple sources within the LABOCA beam (see Weiss et al. 2009).
The corresponding IAU designations are given in Table~\ref{tab:sources}.}
sources found with APEX/LABOCA (Weiss et al. 2009). Of the 126 LESS sources, 17 are located within the GOODS-S field
(Giavalisco et al. 2004). The ALMA observations provide sub-arcsecond location on 9 of the 17 sources located within
the CANDLES coverage of the GOODS-S field, in some cases detecting multiple submillimeter continuum sources within
a single LABOCA footprint. 

In this paper we use the coordinates for the SMGs observed by Hodge et al. (2013) to identify the optical/NIR counterparts
of the SMGs within the CANDELS GOODS-S field. In total we identify and analyze the properties of 10 counterparts. We use
the near-infrared CANDELS data (Grogin et al. 2011; Koekemoer et al. 2011; Guo et al. 2013) as well as optical BViz data
from the GOODS project (Giavalisco et al. 2004), re-measured using PSF matching and the TFIT method (see Guo et al.
2013), and Spitzer/IRAC infrared data (Ashby et al. 2013).
In a recent paper, Targett et al. (2013) also discussed properties of SMGs in the GOODS-S field, using HST/WFC3 data.
In this case, the counterparts were identified using radio continuum as well as Spitzer/IRAC 8$\mu$m data.  Our sample
partly overlaps with that of Targett et al. (2013), with five sources in common. The focus of the two papers are different but
complementary and we compare our results with that of Targett et al. when appropriate. 
We adopt H$_{0}=72$\,km\,s$^{-1}$, $\Omega_{m}=0.3$ and $\Omega_{\Lambda}=0.7$ throughout this paper.
All magnitudes are in the AB system (Oke 1974).

\section{Data and Analysis}\label{data}

\subsection{The Data}
The submm data for the galaxies discussed in this paper are taken from the ALMA 870$\mu$m survey presented in Hodge
et al. (2013). The median angular resolution of the interferometric data is 1\ffas60$\times$1\ffas15. For sources detected at
more than 3$\sigma$, the positional uncertainty of the center of emission is estimated to be 0\ffas2--0\ffas3 (Hodge et al. 2013).
This allows a secure identification of optical and near-infrared counterparts. The entire LESS sample
comprises 126 sources but in this paper we concentrate on the subset of LESS sources that fall within the GOODS-South
field, where an abundance of supporting data are available (Table~\ref{tab:sources}).

We identify optical/near-infrared counterparts using data from the Cosmic Assembly Near-Infrared Dark Energy Legacy Survey
(CANDELS: PIs Faber \& Ferguson; see Grogin et al. 2011 and Koekemoer et al. 2011), with emphasis on the GOODS-S field
(see Guo et al. 2013). CANDELS is a HST Multi-Cycle Treasury Program to image five legacy fields, including the GOODS-South,
with the Wide Field Camera 3 (WFC3) in the near-infrared. For this paper we use the full depth observations over the entire
field. The images were reduced and drizzled to a 0\ffas06 pixel scale to create a full depth mosaic. For further details on
the CANDELS survey strategy and science data pipeline products, please refer to Grogin et al. (2011) and Koekemoer et al. (2011),
respectively.

The photometric data consists of 18 bands ranging from UV to mid-infrared. The data contains UV data (CTIO/MOSAIC and VLT/VIMOS),
optical (HST/ACS F435W, F606W, F775W, F814W, and F850LP), and infrared (HST/WFC3 F098M, F105W, F125W, F140W and F160W;
VLT/ISAAC Ks; VLT/HAWK-I Ks, and Spitzer/IRAC 3.6, 4.5, 5.8, 8.0$\mu$m) observations. See Ashby et al. (2013) for details on the
Spitzer/IRAC data.
A catalog (Guo et al. 2013) was made based on source detection in the WFC3 F160W band. The WFC3 mosaic includes the data from
CANDELS deep and wide observations as well as previous ERS (PI: R. O'Connell; see Windhorst et al. 2011) and HUDF09 (PI: G.
Illingworth; see Bouwes et al. 2011) programs. Photometry of HST bands other than WFC3 F160W is measured using PSF matching.
Photometry of lower resolution images is done using the TFIT template-fitting method (see Laidler et al. 2007; Lee et al. 2012;
Guo et al. 2013). We also make use of the Spitzer/MIPS 24$\mu$m data (Magnelli et al. 2011). The 24$\mu$m flux is not used in the
SED fitting procedure, but is used in estimating the potential influence of a power-law continuum.

The mosaic reaches a 5$\sigma$ limiting depth in the F160W band, within an aperture of radius 0\ffas17, of 27.4, 28.2 and 29.7 AB for CANDELS wide, deep,
and HUDF regions, respectively. The catalog contains 34930 sources with the representative 50\% completeness reaching 25.9, 26.6 and
28.1 AB in the F160W band for the three regions. 
The sources were extracted using SExtractor in both a hot and cold mode. By combining these modes it is possible to define substructures
within a galaxy. For details on source extraction and photometric calibration, see Guo et al. (2013) and references therein. 

There are 17 LESS sources within the GOODS-South field as defined by the Spitzer/IRAC coverage. Of these, 13 fall within the WFC3
F160W observations covered by CANDELS. ALMA data was obtained for 9 of these. Hodge et al. (2013) reported multiple
submm emission regions for 3 of the 9 LESS sources within the WFC3/F160W area, for a total of 14 submm emission sources.
As discussed in Sect.~\ref{id}, we will only retain one of these multiplets as a true individual SMG. Therefore the total  number of
sources analyzed in this paper is 10.

\subsection{Analysis}\label{analysis}
The optical/near-infrared counterparts to the LESS sources were identified by matching the coordinates of the LESS sources
observed with ALMA (Hodge et al. 2013) with the CANDELS WFC3 F160W selected catalog (Guo et al. 2013). Given the positional
accuracy of the LESS data
($\sim0\ffas2-0\ffas3$) and the 0\ffas06 pixel scale of the HST data, we can securely connect the SMGs with their counterparts, with
two caveats. The central location of the submm emission may not coincide with the center of optical/near-infrared emission due to
strong dust extinction. This effect is exacerbated with increasing redshift as shorter wavelength emission is shifted into a
given optical/near-infrared filter. We must therefore allow for coordinate offsets of the order of galactic scales, typically $<1''$, or 8.5\,kpc at $z\sim2$.
The second caveat is the possible, although improbable, coincidence of a background SMG with a foreground galaxy. While this
is unlikely for such a small sample, the fact that the initial LABOCA survey is luminosity limited creates a bias towards sources
that may be gravitationally lensed by a foreground object. We inspected the HST images for signatures of gravitational lensing of
the SMG counterparts, without finding any (Sect.~\ref{results}). 
Note that the lack of lensing signature does not in itself mean that the designated counterpart and the SMG are coincident. If the
redshifts are comparable, they could still be separate objects.
If the separation between the center of the ALMA 870$\mu$m emission and the nearest galaxy exceeds 1\ffas0, we do
not consider the galaxy and the SMG to be associated, unless there is an obvious optical/NIR connection between the submm
emission and the nearest galaxy. This is the case for one of the ALMA LESS sources (LESS 67.2, see Sect.~\ref{id}).

Parameters, such as photometric redshift, stellar mass, age, extinction and metallicity are derived by fitting spectral energy distributions
(SEDs) using the stellar population synthesis models of Bruzual \& Charlot (2003, hereafter BC03) to the counterparts of the LESS sources.
In order to fit the SED of each galaxy in an unbiased and prior-free manner, we explore a large parameter
space for redshift, stellar age, extinction, metallicity, and star formation history. The current version of the SED fitting algorithm is an updated
version of the one described in Wiklind et al (2008). Here we use a Chabrier initial mass function (IMF) with lower and upper mass cutoffs at
0.1 and 100 M$_{\odot}$, respectively (Chabrier 2003). The resulting SEDs are redshifted in the range $z=0.05-10$ with $\Delta z=0.05$. The redshift
resolution is then further refined to $\Delta z=0.01$ using spline fitting in the E$_{B-V}-z$ parameter space. Dust obscuration is introduced
using the attenuation law of Calzetti et al. (2000). It is parametrized through the E$_{B-V}$ color index, covering the range
E$_{B-V}=0.0-1.0$, with $\Delta \mathrm{E}_{B-V}=0.025$. Additional attenuation by neutral hydrogen absorption in the intergalactic
medium (IGM) is added using the Madau (1995) prescription for the  mean IGM opacity.
The age of a stellar population is measured from the onset of star formation. The age range extends from 50 Myr to 7 Gyr, with an
approximate logarithmic step. The maximum age of a galaxy at a given redshift is limited by the age of the universe at that point.
Four different metallicities are used, 0.2, 0.4, 1.0, and 2.5 Z$_{\odot}$.
The star formation history is parametrized as a so-called `delayed-$\tau$' model (see Lee et al. 2010): $\phi(t,\tau) \propto (t/\tau^2)\,\exp(-t/\tau)$.
While having the same number of parameters as the standard exponentially declining $\tau$-models, the delayed-$\tau$ models 
exhibit an increasing star formation rate for $t < \tau$, and a declining part for $t > \tau$. For $t  >> \tau$,
the parametrization resembles the exponentially declining $\tau$ models.
A recent burst of star formation can be parametrized with a young $t$ and a small $\tau$, but for multiple generations of stellar populations
the delay-$\tau$ model, as any other single-component parametrization of the star formation history, will produce an average, luminosity
weighted, spectral energy distribution. In this case, the age and $\tau$ parameters will reflect the average SED.

The stellar mass is estimated from the SED by integrating over the entire spectral energy distribution, giving $L_{bol}$ and then multiplying with the
$M_{*}/L_{bol}$ ratio derived for the best-fit SED. The integration is done over the non-obscured SED, from 92{\AA} to 160$\mu$m, only considering
the stellar SED. This method has been shown to give robust mass estimates when compared with galaxies from semi-analytical models, where the mass
of the stellar component is known (Lee et al. 2010; Wiklind et al. 2008). 

The BC03 models only include stellar components and are not designed to model an AGN contribution to the SED. Since a substantial
fraction of SMGs are known to contain an AGN (e.g. Donley et al. 2010), the SED fitting results could be biased by the AGN's UV contribution. This would be
modeled as a young stellar component and/or as having less extinction present than what is the true case. As discussed in Sect.~\ref{results},
however, this bias does not appear to be present in the SED fitting results for the LESS sources, even though almost 50\% of them
are designated as containing an AGN based on x-ray (Chandra 4Ms survey, Xue et al. 2011), radio (Padovani et al. 2011) and IR signatures
(courtesy J. Donley, see Donley et al. 2013). In fact, the stellar content of the SMGs are characterized by a relatively evolved population.
{This AGN fraction is much higher than the x-ray survey of CDFS SMG's by Wang et al. 2013), who finds that $\sim$10\% of the SMGs exhibit
x-ray emission.}

The possibility that an obscured AGN contributes to the observed fluxes is handled by subtracting a
power-law continuum of the form $f_{\nu} \propto \nu^{-\alpha}$ and evaluating the $\chi^{2}_{\nu}$ for the resulting SED fit.
The power-law index is expected to be in the range $\alpha\sim1-2$ (Caputi 2013). The power-law emission is scaled to the observed IRAC
8$\mu$m flux and the power-law flux is subtracted from the observed photometric values at shorter wavelengths. A new SED fit is done and the
$\chi^{2}_{\nu}$ evaluated. The scaling to the 8$\mu$m flux is done with a multiplicative factor, ranging from 0 (no power-law emission at
8$\mu$m), to 1 (all of the 8$\mu$m emission is due to a power-law component) in steps of 0.05. The SED fit with the minimum $\chi^{2}_{\nu}$
is selected as the best-fit SED. This approach is similar to that of Hainline et al. (2011), Michalowski et al. (2012) and Caputi (2013).

The slope of the power-law emission varies depending on the type of AGN and/or star formation activity that is the source of the emission.
Nearby AGNs of type-1 have slopes $\alpha\sim1-1.5$ (Edelson \& Malkan 1986; Neugebauer et al. 1987) and local Seyfert 2 galaxies
are characterized by a steeper continuum SED with $\alpha\sim2-4$ (Alonso-Herrero et al. 1997). Obscured starburst galaxies will
have a near- to mid-infrared slope that is similar to that of type-2 AGNs. The average mid-infrared slope for SMGs determined from
Spitzer IRS data is $\alpha=1.97 \pm 0.22$ (Men\'endez-Delmestre et al. 2009). The latter sample contains SMGs similar to the
ones discussed here, and we will assume $\alpha=2.0$.
A steeper slope would have a minimal impact on the observed fluxes. The slope could, however, be shallower.
When extrapolated to 24$\mu$m, the power-law continuum flux should not exceed the observed 24$\mu$m flux. We will therefore use the
24$\mu$m flux to constrain the slope of the power-law continuum, keeping in mind that for some redshifts the 24$\mu$m flux may
get contribution from PAH features. Constraining the value of $\alpha$ leads to a power-law flat enough to subtract more than the
observed emission at short wavelengths. This may be an artifact of extending the power-law into restframe UV wavelengths.
We therefore impose an arbitrary limit to the subtraction of the continuum to only be applied to fluxes at restframe $\lambda>8000$\AA.
Except for the cases of a very flat power-law ($\alpha\la1$) this limitation does not affect the outcome of the fitting and subtraction process.

In order to assess the stability of the best fit SED with respect to photometric uncertainties, we did a Monte Carlo simulation where the
photometric values are allowed to vary stochastically within their nominal errors. The flux errors are assumed to normally distributed and
uncorrelated (see Wiklind et al. 2008 for a more detailed description of the procedure). We generate $10^{3}$ photometric data sets for
each galaxy. Bands with non-detections are still treated as upper limits. We then determine the best-fit parameters for each realization
of the photometric data in the same manner as described above. The resulting distribution of the best-fit values for each parameter
represents the probability distribution for this particular parameter and allows an estimate of the confidence of the various solutions
with respect to the photometric values and uncertainties. The outcome from the Monte Carlo simulations is often a well-behaved (e.g.
close to Gaussian) distribution of values for the parameters (in particular photometric redshift and stellar mass). There are, however,
cases where the distribution is strongly non-Gaussian and where the solution `flips' between very different parameter values where the
solutions can have equal weight in terms of their $\chi^{2}$ values. This behavior can be due to aliasing between the Lyman and Balmer breaks
(e.g. Dahlen et al. 2013). 

For each solution of a photometric redshift we construct a probability distribution $P(z) \propto \exp(-\chi^{2})$. This distribution
represent the uncertainty of the photometric redshift associated with the fitting of a single set of photometric values. Combining
the $P(z)$ distributions from the Monte Carlo simulations gives a probability distribution that takes the stability of the solutions
into consideration. We combine the $P(z)$'s  by taking the median value for each redshift bin $\Delta z=0.01$. Since the 
resulting $P(z)$ can still exhibit multiple peaks, we assigned the peak value as the most probable photometric redshift. 

The photometric redshifts obtained in the SED fitting process with this code has been tested together with 11 other codes (see
Dahlen et al. 2013) and found to give results consistent with both a spectroscopic comparison sample and the other codes.

\subsection{Morphological parameters}\label{morphpar}
We measure the CAS (concentration, asymmetry and clumpiness) parameters of the SMG counterparts using HST/WFC3 F160W
images. The morphology of the counterparts may reveal clues to their origin and evolution. The  CAS parameters  are a non-parametric
method for measuring the forms and structures of galaxies in resolved CCD images (for further details see Conselice et al. 2000;
Bershady et al. 2000;   Conselice 2003). We also employ visual inspection of the counterparts and their immediate environment
to identify signs of gravitational interaction.

The asymmetry of a galaxy is measured by taking the image of a galaxy and rotating it 180$^{\circ}$ about its center, and then
subtracting these two images (Conselice et al. 2000). There are corrections done for background, and radius (see Conselice et al.
2000). The center for rotation is decided by an iterative process which finds the location of the minimum asymmetry.
We also derive the Gini and M$_{20}$ coefficients for the SMGs. The Gini parameter is a statistical tool originally used to determine
the distribution of wealth within a population, with higher values indicating a very unequal distribution. The M$_{20}$ parameter is
similar to the concentration parameter in that it gives a value that indicates if light is concentrated within an image; it is, however,
calculated slightly differently from C and Gini (e.g. Lotz et al. 2004, 2008).

We compare the non-parametric morphology parameters of our SMGs with those of local galaxies as well as a sample of $z\sim2$
Lyman break galaxies (Sect.~\ref{morph}), but the final designation of morphology class is done by visual inspection of the
high resolution HST images. We designate a SMG counterpart as a merger if it shows signs of gravitational interaction and has tidal
tails. If the counterpart has one or more neighbors within 30\,kpc and $\Delta z_{\mathrm{phot}}\pm$0.2, but without signs of interaction,
we designate it as `neighbor'. In the absence of any interaction and neighbor within 30\,kpc, the SMG is designated as `isolated' (see
Sect.~\ref{morphdiscussion}).

\section{Results}\label{results}

\subsection{Identification of the counterparts}\label{id}
The LESS sources that fall within the CANDELS coverage of the GOODS-South field, and for which ALMA data were obtained, are listed
in Table~\ref{tab:sources}, together with the nearest galaxy as defined in Sect.~\ref{analysis}.
Three of the LESS sources detected with LABOCA are reported as multiple submm sources in Hodge et al.
(2013), LESS 17, LESS 67 and LESS 79, resulting in a total of 14 sources. We only retain one of these multiplets in our analysis of
the SMG counterparts. We describe these cases below:

{\bf LESS 17}: The ALMA observation of LESS 17 shows one main 870$\mu$m emission source within the ALMA primary beam, and two
fainter sources outside the primary beam\footnote{The primary beam is defined as the diameter where the sensitivity is 50\% of that in the
beam center. For the ALMA 870$\mu$m observations, the diameter of the primary beam is 18'', similar to the original LABOCA footprint.}
(LESS 17.2 and LESS 17.3). The fainter sources are part of the 'Supplementary sample' of Hodge et al. (2013) because of their location
outside the primary beam. The 870$\mu$m fluxes of LESS 17.2 and LESS 17.3, after correcting for the primary beam response, are
3.7$\pm$0.9 and 5.1$\pm$1.2 mJy, respectively, but they both have a flux correction factor $>$2. As noted by Hodge et al. (2013),
the number of spurious
detections increases for the area outside the primary beam because of a rapidly increasing noise level. There are no sources in the
CANDELS images within 2\ffas3 of LESS 17.2. For LESS 17.3 there is a CANDELS source at a distance of 0\ffas8, (CANDELS ID 28579)
with  $z_{\mathrm{phot}}=2.35$ and a stellar mass of 7$\times$10$^{8}$\,M$_{\odot}$.
We re-analyzed the original ALMA data and were not able to reproduce the LESS 17.2 and LESS 17.3 detections. We did find several
$\sim$2-3$\sigma$ peaks outside the ALMA primary beam, but not at the positions of LESS 17.2 and LESS 17.3 as reported by Hodge
et al. (2013). We will therefore not include LESS 17.2 and LESS 17.3 in our analysis and refer to the main submm source as LESS 17.

{\bf LESS 67}: Hodge et al. (2013) report two submm emission regions associated with this source, LESS 67.1 and LESS 67.2, both within
the ALMA primary beam and both part of the `Main sample' of Hodge et al. (2013). As we discuss below, these two sources appear to be
associated with a single merging system. We will therefore treat these two submm sources as a having a single counterpart and refer to
it as LESS 67.

{\bf LESS 79}: There are three submm emission sources reported for this source, LESS 79.1, LESS 79.2 and LESS 79.4, all within
the ALMA primary beam. While LESS 79.1 and LESS 79.2 are directly associated with CANDELS counterparts, the closest neighbor to
LESS 79.4 is more than 2'' away (a low-mass galaxy 9$\times$10$^{8}$\,M$_{\odot}$, CANDELS ID 26896). The 870$\mu$m flux of LESS 79.4 is
only 1.8$\pm$0.5 mJy but it is classified as part of the 'Main sample' in Hodge et al. (2013). Based on the low flux level and the lack of a
counterpart in the CANDELS images, we will not include LESS 79.4 in the following discussion.

\begin{figure}
\epsscale{0.9}
\plotone{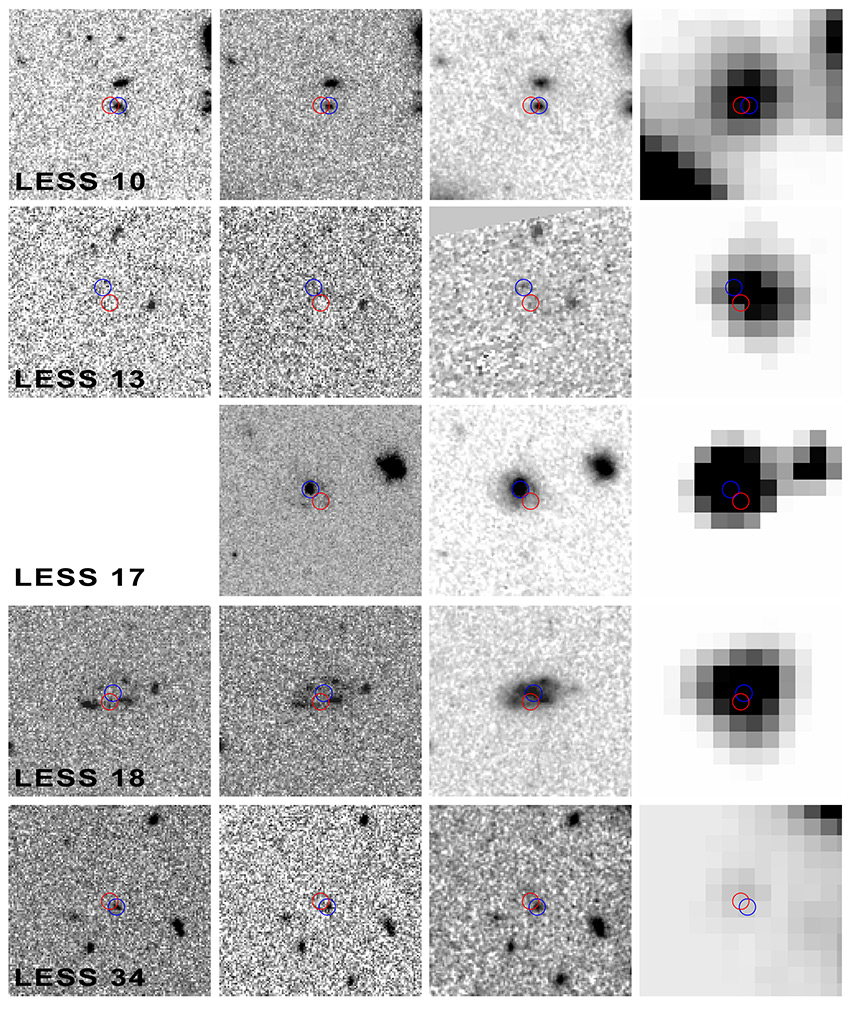}
\caption{Cut-outs for LESS sources observed and detected with ALMA (Hodge et al. 2013). Each image is $7''\times7''$ in size and
are from left to right: ACS/F606W, ACS/F850LP, WFC3/F160W and IRAC/3.6$\mu$m (Spitzer/IRAC images from Ashby et al. 2013). 
The circles have a radius of 0\ffas3 and mark the center position for the ALMA data (red) and the WFC3/F160W (blue).
For LESS 67 a second red circle marks the position of LESS 76.2, likely to be a part of the tidal arm.} 
\label{fig:composite1}
\end{figure}

\begin{figure}
\epsscale{0.9}
\plotone{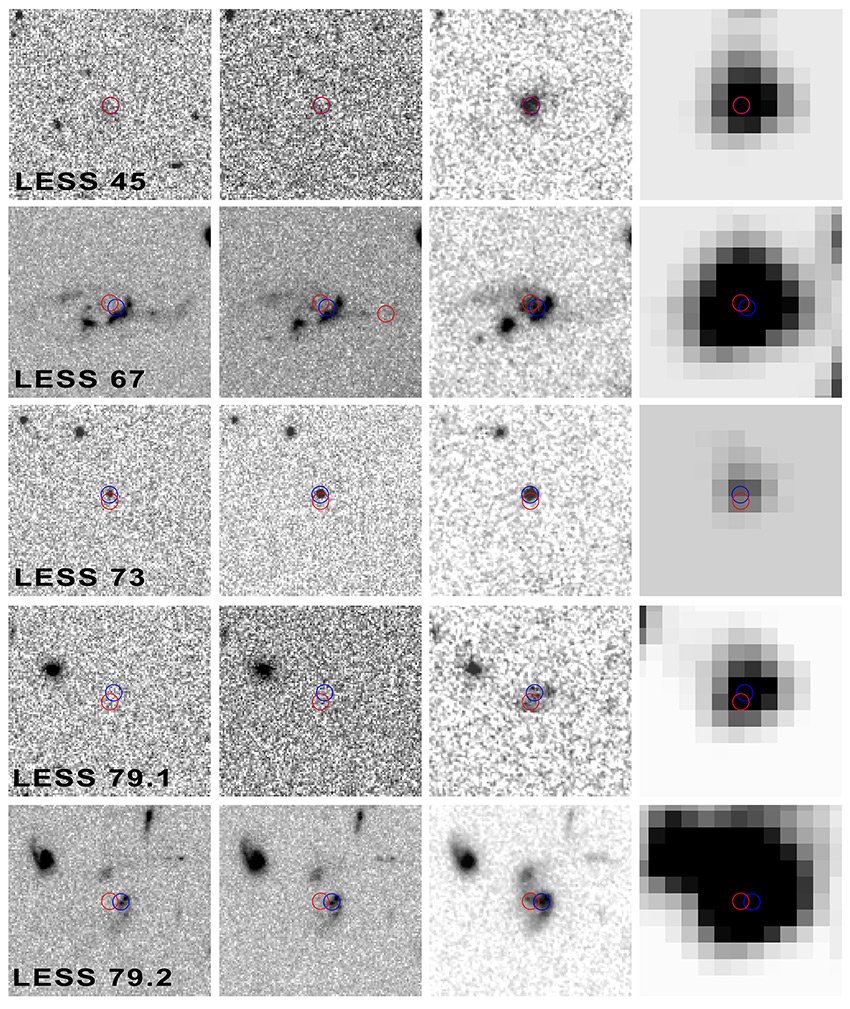}
\caption{Same as Fig~\ref{fig:composite1}.} 
\label{fig:composite2}
\end{figure}

\begin{figure}
\epsscale{0.9}
\plotone{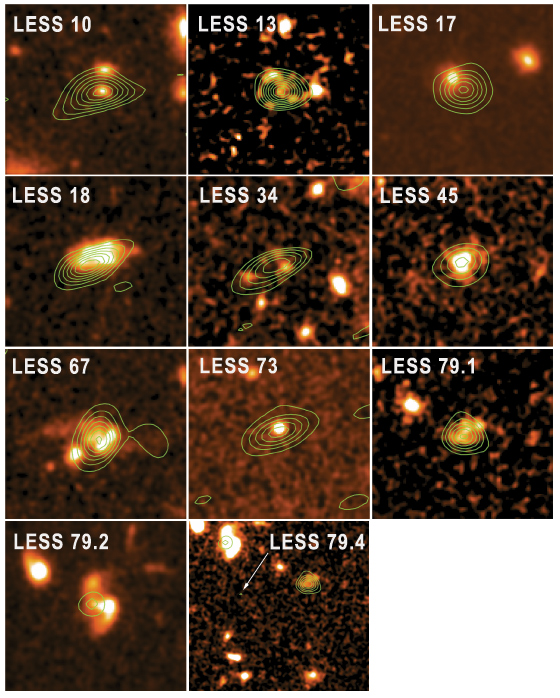}
\caption{Smoothed versions of the WFC3/F160W images of the SMGs with ALMA submm emission contours overlayed. The lowest contour is 1.0\,mJy/beam for each SMG, except for LESS 67, where it is 0.5\,mJy/beam (in order to see LESS 67.2). The contour steps are 0.5\,mJy/beam for all sources except LESS 45 and LESS 73, which have steps of 1.0\,mJy/beam. The images are $7''\times7''$ in size and smoothed using a 3-pixel Gaussian. The last frame has a slightly larger scale and shows LESS 79.1, LESS 79.2 and LESS 79.4 (indicated by an arrow).}
\label{fig:smooth}
\end{figure}

The final number of SMGs in our sample is thus 10.  Images of the area covering the location of the 10 ALMA LESS source are given
in Figs.~\ref{fig:composite1}~\&~\ref{fig:composite2}.  In Fig.~\ref{fig:smooth} we show WFC3/F160W images of the SMGs, smoothed with a 3-pixel
kernel Gaussian, with the ALMA 870$\mu$m contours superimposed. Each image shown in Figs.~\ref{fig:composite1}$-$\ref{fig:smooth} is $7''\times7''$
in size and centered on the coordinates of the LESS sources derived from the ALMA data. These coordinates are listed in Table~\ref{tab:sources}
together with the separation between each LESS source and the center of the nearest galaxy in the WFC3 F160W image. The nearest galaxy
within a radius of 1'' is designated as the counterpart (see definition in Sect.~\ref{analysis}).

The average separation between the center of the ALMA LESS and WFC3/F160W sources is 0\ffas36, corresponding to $\sim3$ kpc at $z=2$.
The average separation is similar to the accuracy of the center positions derived from the ALMA data (see Sect.~\ref{data}) and it is therefore
not clear whether offsets of $\sim$0\ffas3 are real or simply due to positional uncertainties.
The ALMA positions are defined in the International Celestial Reference Frame (ICRF), based on distant QSOs, while the CANDELS positional
data is defined in the FK5 reference frame. These two frames have a relative positional uncertainty less than $\pm$50\,mas (e.g. Arias et al. 1995 and
references therein).

Three of the SMGs show slightly larger offsets from their optical/NIR counterpart, with offsets in the range 0\ffas5-0\ffas6 (LESS 13, 17 and 79.2).
LESS 79.2 has a distorted morphology and is a merger according to the definition in Sect.~\ref{morphpar}. The offset between the center of the
optical and submm emission regions is $\sim$4\,kpc, similar to that  seen in the Antennae galaxy (Wilson et al. 2000).
The submm flux of LESS 79.2 is, however, only 2.0$\pm$0.4\,mJy, the lowest in our SMG sample. This leads to a slightly worse
positional accuracy of the submm flux compared with stronger sources. LESS 13 appears to be associated with a single source in
Fig.~\ref{fig:composite1}, but the smoothed and stretched version (Fig.~\ref{fig:smooth}; see Sect.~\ref{morph}) reveals that it is really part of a
merging system. The second component has the same photometric redshift and stellar mass as the designated counterpart. The counterpart of
LESS 17, on the other hand, does not show any signs of a close companion or distorted morphology. It is among the stronger submm sources
in our sample and the offset of 0\ffas59 ($\sim$5 kpc) between the center of optical and submm emission is likely to be real.

Before deriving properties of the counterparts to the ALMA LESS sources, we need to discuss LESS 34. This source is part of the `Supplementary
sample' of Hodge et al. (2013) and is thus viewed as less robust than those in the `Primary sample'.  LESS 34 is detected inside the ALMA primary
beam (close to the center) and it has a flux of $S_{850\mu m}=4.5\pm0.6$, giving a S/N$=$7.7. The noise level of the ALMA map is as good as any
of the LESS sources in the `Primary sample' with $\sim$0.55\,mJy/beam. The only reason this source is part of the `Supplementary sample' is that
the ratio of the major and minor axes of the ALMA interferometric beam is 2.3. One of the criteria used by Hodge et al. (2013) to distinguish between
the `Primary' and `Supplementary' sample is that the ratio of the major and minor axes should be $<$2. LESS 34 was observed at slightly low elevation,
causing the interferometric beam to be more elongated than what is the case for some of the other LESS sources. We re-imaged and re-analyzed the
ALMA results for LESS 34 and achieved the same result as in Hodge et al. (2013). We tried different weighting schemes and CLEAN parameters, but
the source remains a robust detection in all cases. We will therefore retain this ALMA LESS source in our analysis.

Seven of the SMGs in our sample are also part of the sample in Targett et al. (2013): LESS 10, 18, 34, 45, 67, 73, and 79.2. Overall, we find excellent
agreement with the Targett et al. identifications, with 5 of the 7 LESS sources identified with the same counterpart in the two studies. The power of the
ALMA data is highlighted in three of these seven cases, where we are able to identify robust counterparts to the potential multiple (or single
mis-identified) counterparts in Targett et al (2013).
Firstly, Targett et al. identified LESS 34 with a z$_{spec}=$1.08 spiral galaxy, 5\ffas3 away from the center of the submm emission seen with ALMA.
The actual counterpart is a faint and otherwise unremarkable galaxy. LESS 34 is not radio detected and Targett et al. based their
identification solely on 8$\mu$m emission. This association was therefore less secure.
In the case of LESS 10, Targett et al report the identification as a close galaxy pair separated by 0\ffas8, but could not differentiate between them.
Using the ALMA data, we are able to securely associate the submm emission with the southern counterpart. The two galaxies making up the pair
have different photometric redshifts and appear to be a chance superposition.
Finally, LESS 79 possess two submm emission regions within the original LABOCA footprint. Based on the 8$\mu$m emission, Targett et al. 
selected 79.2 as the counterpart, but the ALMA data shows that LESS 79.1 is the stronger of the two submm emission regions.
This comparison of single-dish radio+8$\mu$m identification from Targett et al to our ALMA based selection highlights the strengths of both methods,
and the obvious benefit of ALMA data to remove the confusion associated with the large beam sizes of previous submm instruments. Photometric
redshifts and stellar mass estimates for the sources with identical counterparts in both samples are shown in Table~\ref{tab:targett}.

\subsection{Photometric redshift and stellar properties}\label{photoz}
\begin{figure}
\epsscale{0.75}
\plotone{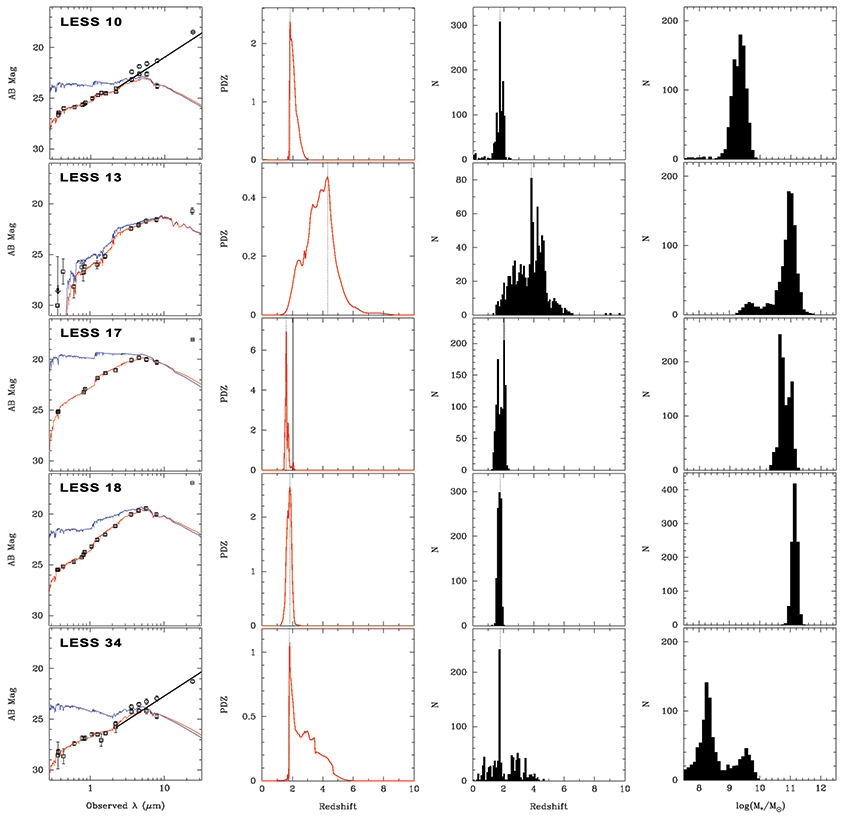}
\caption{{\bf First panel}:\ SED fit results for the LESS SMGs. The red line corresponds to the best-fit SED with dust extinction. The blue line is the same SED corrected for dust . The observed data points are marked, including MIPS 24$\mu$m. The latter is not used in the SED fit. Photometric values corrected for
the presence of a power-law continuum when applicable (see text). For LESS 10 and LESS 34 we also show the photometric values before the
power-law subtraction as well as the power-law continuum. {\bf Second panel}:\ The probability density of the photometric redshift from the SED fit.
The dashed vertical line corresponds to the peak of the distribution. For those SMG's with spectroscopic redshift, $z_{\mathrm{spec}}$ is indicated 
with a full-drawn vertical line. {\bf Third panel}:\ Results from the Monte Carlo simulations of the photometric redshift (see text for details). For those
SMG's with spectroscopic redshift, $z_{\mathrm{spec}}$ is indicated with a full-drawn vertical line. {\bf Forth panel}:\ Results from Monte Carlo simulations
for the stellar masses.} 
\label{fig:sed_a}
\end{figure}

\begin{figure}
\epsscale{0.75}
\plotone{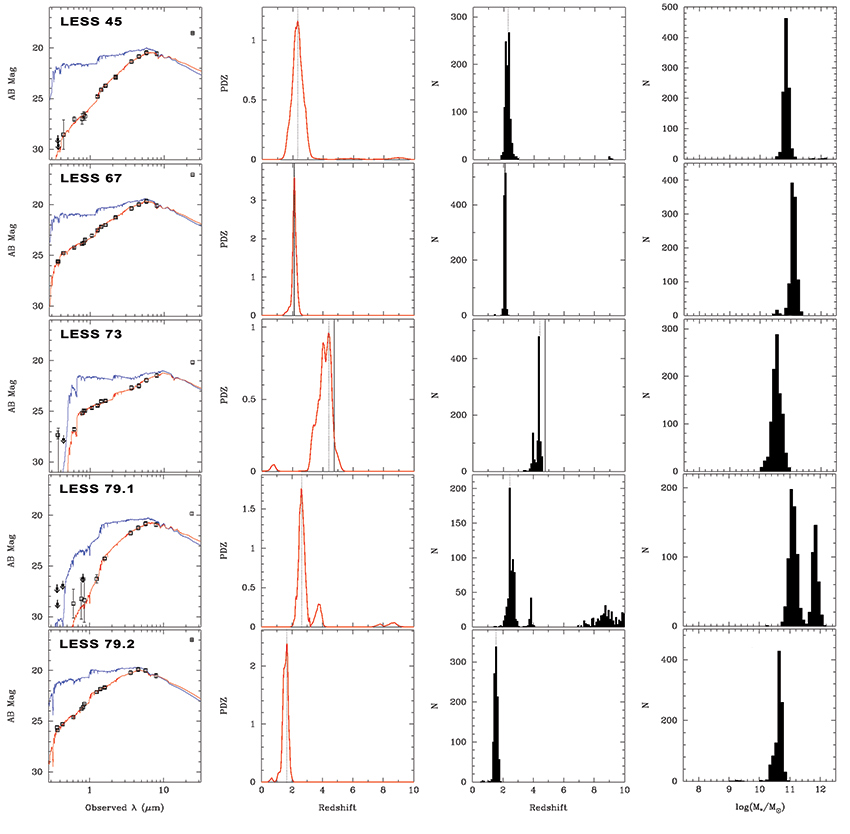}
\caption{Same as for Figure~\ref{fig:sed_a}.} 
\label{fig:sed_b}
\end{figure}

We derive photometric redshifts for all the SMG counterparts. The photometric redshifts are listed in
Table~\ref{tab:sources}. In Figs.~\ref{fig:sed_a} and \ref{fig:sed_b} we show the best-fit SED fits, the resulting photometric redshift probability
distribution, $P(z)$, based on the $\chi^{2}_{\nu}$ values, and the distribution of photometric redshifts obtained from the Monte Carlo simulations
(Sect.~\ref{data}). The Monte Carlo results represents the confidence in the derived redshifts based on the photometric uncertainty, while the
$P(z)$ distribution represents the uncertainty for the photometric redshift based on a single set of photometric values. In general, these two
distributions are similar, but there are a few exceptions; LESS 13 shows a rather broad $P(z)$ distribution, but the Monte Carlo simulations
favor the peak in $P(z)$ around $z_{\mathrm{phot}} =4.20$. The $P(z)$ of LESS 34 shows a fairly high probability of a photometric redshift
in the range z$=1.8-4.5$, with a peak at $z_{\mathrm{phot}}=1.81$. The Monte Carlo simulations clearly favor the peak seen in the $P(z)$
distribution. 

Spectroscopic redshifts are available for 4 of the galaxies (see Table~\ref{tab:sources}). The spectroscopy is of lower quality for two galaxies
(LESS 17 and LESS 79.2). While the photometric and spectroscopic redshifts agree well for LESS 17, the photometric redshift for LESS 79.2
gives a lower value than the spectroscopic redshift. Since both the $P(z)$ distribution and the Monte Carlo strongly favor the lower redshift
solution over the spectroscopic value, we adopt the photometric result as the most likely redshift. For LESS 67 and LESS 73 the spectroscopic
and photometric redshifts agree quite well, and we adopt the spectroscopic values for the SED fit. Note that LESS 73 corresponds to the source
previously identified as a $z=4.76$ SMG by Coppin et al. (2009). The final redshift values are listed in Table~\ref{tab:results},
For the merger systems (LESS 13, 67 and 79.2) we also include results for the secondary galaxy (`B' in Table~\ref{tab:results}). The primary
galaxy (`A') is the one closest to the submm emission and is the designated counterpart.

The photometric values used in the fitting process have been corrected for a near- to mid-infrared power-law continuum for four of the sources
(see Sect.~\ref{data}). The impacted SMGs are LESS 10, 18, 34 and 45\footnote{We also subtracted a power-law continuum from the merger
component `B' in LESS 13, resulting in a small corrections to the derived parameters.}. The effect on the photometric redshifts is negligible.
LESS 18 and 45 have a very small fraction of their 8$\mu$m flux attributed to a power-law continuum (20\%) and the impact on the $\chi^{2}$
value and the resulting parameters is marginal. For LESS 10 and 34, however, the best fit of the SED is achieved if 80$-$90\% of the observed
8$\mu$m flux is due to a power-law continuum and the goodness-of-fit is significantly improved after removing the power-law component. LESS 10
has an apparent neighbor 0\ffas8 to the north (Fig.~\ref{fig:composite1}), but the two sources show no signs of interaction. The northern component
has a somewhat uncertain photometric redshift of $z_{phot} \approx 2.6-3.8$ while the source associated with the 870$\mu$m emission has a fairly
secure redshift of $z_{phot}=1.85$. The two sources could be blended in the IRAC and MIPS images. The IRAC fluxes of LESS 10 are abnormally
high compared to the observed optical and near-infrared fluxes and the SED fit is not able to find a parameter combination that can fit both the
ACS/WFC3 and IRAC data simultaneously. The second source with an apparent strong power-law continuum is LESS 34. In this case there is no
neighbor that can contribute to the IRAC and MIPS fluxes. LESS 34 has unusually faint IRAC fluxes even before subtracting the PL continuum, with
$f_{8\mu m} = 3.0$\,$\mu$Jy,  almost 4 times fainter than any of the other SMGs. The 8$\mu$m fraction ($\delta f_{8}$) and the power-law index
($\alpha$) used in the subtraction of a power-law continuum are listed in Table~\ref{tab:results}. The 24$\mu$m fluxes used in the power-law
estimation are listed in Table~\ref{tab:sources}. The slope of the power-law continuum is consistent with $\alpha\sim2$.  As described in
Sect.~\ref{data}, the slope was not formally part of the fitting process. It was set to  $\alpha=2.0$ and only allowed to decrease if the extrapolated
flux at 24$\mu$m exceeded the observed 24$\mu$m flux. 

The stellar masses of the SMG counterparts range from $M_{*} = 2 \times 10^{8}-1.6 \times 10^{11}$\,M$_{\odot}$, with an average
$M_{*} = (8.5 \pm 6) \times 10^{10}$\,M$_{\odot}$ and median $M_{*} = 9.1 \times 10^{10}$\,M$_{\odot}$.
The two galaxies with significant power-law continuum, LESS 10 and 34, have lower stellar masses by almost two orders of magnitude compared
with the rest of our sample. While both are intrinsically faint at restframe UV through near-infrared wavelengths, their
submm fluxes are typical of the LABOCA detections in the whole ECDFS. The low masses are not due to the subtraction of a power-law continuum.
In the case of LESS 10, the stellar mass is actually higher by 0.2 dex after subtraction of the power-law continuum, while the mass for LESS 34 is
0.04 dex lower after subtraction. The reason for the slightly higher mass for LESS 10 is due to the age parameter. Without the subtraction, the best
SED fit indicates a much younger age, resulting in a lower $M/L$ ratio and, hence, a lower stellar mass. The age of LESS 34 is 50 Myr both with
and without a continuum subtraction. 
We also did the SED fit of LESS 10 and LESS 34 without considering the IRAC data. This leads to the same result as the case with the
IRAC data, but with improved $\chi^{2}_{\nu}$ values. This shows that the SED cannot simultaneously fit the optical/NIR and IRAC data.
Disregarding LESS 10 and LESS 34, the average stellar mass is $M_{*} = (1.1 \pm 0.4) \times 10^{11}$\,M$_{\odot}$ with a median
$M_{*} = 9.6 \times10^{10}$\,M$_{\odot}$ While quite massive, these numbers are less than some of the early estimates (e.g. Borys et al. 2005)
and more similar to those of Hainline et al. (2011), Michalowski et al. (2012) and Magnelli et al. (2012).
Including the second component for the merger systems does not change these results significantly.
The only merger with similar masses for the two components is LESS 13, where the mass ratio is 1.6. For LESS 67 and LESS 79.2 the merger
mass ratio is 20 and 5, respectively (Table~\ref{tab:results}).

Using a different set of SED templates, such as the Maraston et al. (2005) models, with a different treatment of thermally pulsating AGB stars,
would lead to a lower stellar mass by a factor $\sim$2. Also, our SED fitting procedure does not include nebular lines.
The presence of strong nebular lines could enhance the Balmer break for certain combinations of observed bands and redshifts.
This would artificially increase the estimated age when using templates only including a stellar component, leading to an overestimate of the
stellar masses. For the adopted photometric redshifts, the H$\alpha$ line could potentially enhance the Balmer break for LESS  13 and 45 (H$\alpha$
in the K-band) and for LESS 73 (H$\alpha$ in the IRAC 3.6$\mu$m band). However, none of these three SMGs exhibit extraordinarily old
ages or high stellar masses. Hence, at least for the SMG's presented in this paper, the presence of nebular lines is likely to have a minor impact
on the derived parameters.

The age of the stellar population, derived from the SED fits, ranges from 50 Myr (LESS 34) to 2-3 Gyr (LESS 18 and 67). The average/median
age is 0.9/0.7 Gyr. The age parameter depends on the adopted parametrization of the star formation history. In the case of the delay-$\tau$
models, an alternative measure of the stage of the star formation is given by the ratio $t/\tau$, where $t$ is the age of the stellar population.
 For $t/\tau<1$ the star formation rate increases with time, for $t/\tau>1$ the star formation rate decreases with time,
and $t/\tau=1$ indicates that the star formation rate has reached a peak and changes relatively little with time. 
Three of the SMGs show increasing star formation rates; including the least massive galaxies (LESS 10 and LESS 34) and the merger system
LESS 79.2. Five of the SMGs have $t/\tau>2$ and two SMGs have $t/\tau\approx1$. This suggests that $\sim$70\% of the SMGs are in
a fairly advanced stage of the build-up of stellar mass. That two of the least massive galaxies have $t/\tau < 1$, suggests that they may
be in an earlier stage of the SMG phase.
It is important to keep in mind that the range of $t/\tau$ values is limited by the range of parameters used in the SED fitting process:
$t=0.05-7$\,Gyr and $\tau=0.01-2$\,Gyr. This means that a galaxy with an indicated age of 2 Gyr or older is limited to a $t/\tau \ge 1$. 

The implied  dust extinction values, parametrized through their E$_{B-V}$ values are very high for all of the SMGs in our sample. Metallicities
are also high, with solar or super-solar being the norm. Only LESS 17 has a value suggesting less than solar metallicity. It should be noted that
we only use 4 different metallicities in the SED fit and that there is a degeneracy between metallicity and age, as well as dust extinction. These
values are therefore less robust then those for photometric redshift and stellar mass.

\subsection{Morphology and environment}\label{morph}
The CAS values together with the M$_{20}$ and Gini coefficients for the SMGs are listed in Table~\ref{tab:morph}. 
We will not discuss the concentration and clumpiness parameter ('C' and 'S'), but include them in Table~\ref{tab:morph}
for completeness. We just note that the concentration parameters for the LESS sources are typical for massive galaxy profiles
(e.g. Conselice et al. 2003).
A value of the asymmetry index, $A>0.35$, is usually taken as an indication of a merging system (Conselice et al. 2003). The asymmetry
index for LESS 34, 67 and 79.2 is greater than 0.35 and the latter two sources are visually identified as mergers based on the presence
of tidal tails. The merger LESS 13, however, has the smallest asymmetry parameter of our sample, $A=0.09$, but it is also very faint and
the morphological characteristics only emerges after smoothing the original image.

In order to assess whether the asymmetry values of the SMGs are different from non-SMG galaxies of similar mass and at the same
redshift, we constructed a control sample for each SMG, consisting of galaxies with stellar mass within $\pm$0.2 dex
of that of the SMG and $z_{\mathrm{phot}}$ within $\pm$0.1 of the photometric redshift of the SMG in question. The control galaxies were
drawn from the CANDELS GOODS-S F160W selected catalog. The number of galaxies in each of the 10 control samples is in the range 20-2000,
depending on the redshift of the SMG and its stellar mass.
We then derived the $A$ parameter for the galaxies in the control samples. Comparing this $A$ value with the SMG's shows that the
three SMGs with the highest asymmetry parameter (LESS 34, 67 and 79.2) have $A$ values that are higher than 98\% of their respective control
sample. The $A$ value for LESS 79.1, the 4th largest among the SMGs, is higher than 80\% of its control sample.

We also did a Monte Carlo simulation were we compared the average $A$ value of our SMGs with that of a randomized sample of galaxies.
The galaxy samples were constructed by randomly selecting one galaxy from each of the control samples defined for the individual SMGs
(see above), resulting in 10 randomized galaxies. The average $A$ value of these 10 galaxies was derived and compared with the average
$A$ value of the SMGs. Repeating this procedure $10^{3}$ times, we find that in 99\% of the Monte Carlo realizations, the average asymmetry
of the SMG counterparts is higher than the average $A$ value of the control galaxies. This suggests that the SMGs are intrinsically more asymmetric
than the typical galaxies at the same redshift and mass range.

\begin{figure}
\epsscale{0.95}
\plotone{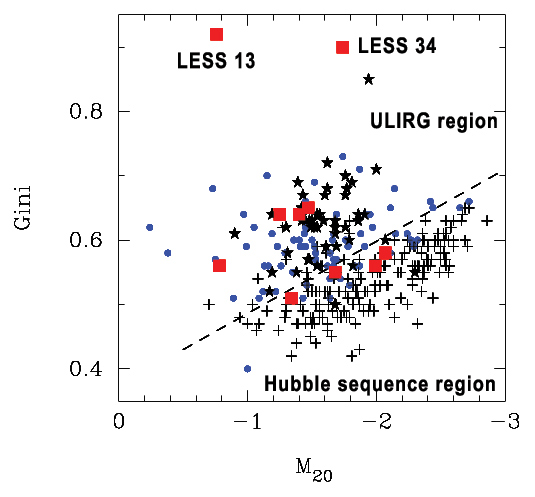}
\caption{The Gini and M$_{20}$ indices for the SMGs (red squares) plotted together with a sample of local ULIRGs (blue
circles), local E-Irr galaxies (crosses) and a sample of z$\sim$2 Lyman break galaxies (stars). The dashed line delineates
the region of local ULIRGs and z$\sim$0 Hubble sequence galaxies (Lotz et al. 2004). Four of the SMG's fall in the latter
region: LESS 10, 17, 18 and 73. } 
\label{fig:gini}
\end{figure}

In Fig.~\ref{fig:gini} we compare the Gini and M$_{20}$ indices for our SMGs with a sample of local galaxies, including ULIRGs as well as
normal early-type, spiral and irregular galaxies. The comparison samples are the same as in Conselice (2003; see Table 1, 4 and 5) and
Lotz et al. (2004). From Lotz et al. (2004) we also used a sample of z$\sim$2 Lyman break galaxies (LBG). For local galaxies, Lotz et al.
(2004) found that the ULIRGs and early-type/disk/irregular (now called `normal') galaxies occupy distinct regions in the
Gini-M$_{20}$ plot. This division is shown with a dashed line in Fig.~\ref{fig:gini}. Six of the SMGs are located in the region occupied by
local ULIRGs. The four SMGs that fall in the `normal' galaxy region (LESS 10, 17, 18 and 73) are all close to the dividing line between the two
regions. At first sight this suggests that the SMGs are similar to local ULIRGs, which are almost exclusively major mergers. However, the z$\sim$2
LBGs, which are characterized by extended disk structures and not dominated by major mergers, also occupy the same region of the M$_{20}$-Gini
plane as local ULIRGs and SMGs, suggesting that these indices may not be applicable for high redshift objects in the same manner as for low
redshift galaxies.

Visual inspection of the SMG counterparts shows that LESS 67 and LESS 79.2 are mergers. Both of these have a large asymmetry
index. Smoothing the WFC3/F160W images with a Gaussian (FWHM=3 pixels) and applying a hard stretch reveals that LESS 13 is also a
merger system with two galaxies and tidal tails. A third galaxy, to the north, has a photometric redshift $z_{\mathrm{phot}}\approx 2.6$, and
is not related to the submm emission. The smoothed image of LESS 34 does not reveal any obvious signs of interaction or merger
(Fig.~\ref{fig:smooth}). A faint object appears $\sim$1\ffas0 away from LESS 34, but no photometric redshift can be obtained for this
galaxy and it is not clear if it is related to LESS 34. Based on the lack of any signs of tidal features, we do not consider LESS 34 as a merger,
despite its high $A$ value. There are two galaxies within $\sim$30\,kpc that have photometric redshifts compatible with the counterpart of
LESS 34, but without any signs of gravitational interaction. 

LESS 17 has the largest offset between the center of the submm emission and the designated counterpart (Table~\ref{tab:sources}).
The smoothed image, however, does not reveal any indication of a neighbor, nor any distortion to the designated counterpart. An
equally bright galaxy $\sim$3\ffas0 away has a photometric redshift incompatible with the SMG counterpart. In this case, the submm emission
could be associated with a close neighbor completely obscured at restframe optical wavelengths. The lack of any distortion of the
designated counterpart suggests that the submm emission could be associated with a background galaxy. Only spectroscopic
data of the counterpart and the SMG itself will be able to determine whether the the optical and submm emissions are physically
associated.
 
LESS 79.1 is very faint but in the smoothed image it is shown to be a clumpy galaxy. In this case, the asymmetry parameter is the 4th highest
of the SMG sample and in this case indicates a clumpy galaxy rather than an on-going merger. Based on $z_{\mathrm{phot}}$, the bright galaxy
seen $\sim$2\ffas5 away from LESS 79.1 in (Fig.~\ref{fig:smooth}) is not associated with the SMG. 

From the visual inspection, and using photometric redshifts, we find 3 SMGs that are part of mergers (LESS 13, 67 and 79.2), 2 SMGs with
neighbors within 30\,kpc (LESS 34 and 45), and 5 that appears to be completely isolated (LESS 10, 17, 18, 73 and 79.1); see Table~\ref{tab:results}.
Three of the SMGs deignated as isolated or with neighbors, are associated with a clumpy counterpart (LESS 18, 45 and 79.1).
We emphasize that the determination whether or not a galaxy is a merger rests on the appearance of tidal tails and obvious near companion(s).
The lack of these features does not necessarily preclude that a given galaxy is a merger, perhaps in a later evolutionary stage. Attempting to
interpret the SMGs in an early and late stage (see Sect.~\ref{nature}) does not, however, support the notion of the mergers being an early stage
of the SMG phase.

\subsection{Star formation and gas/dust content}\label{gas}
The molecular gas mass of SMGs is usually measured using one or more CO rotational transitions. With the exception of LESS 73, for which
Coppin et al. (2010) observed the CO(2-1) line and derived a molecular gas mass of $1.6\times10^{10}$\,M$_{\odot}$, no CO observations
have been done of the SMGs in our sample. The dust mass, however, can be estimated from the submm flux. As shown by Scoville (2012) and
Scoville et al. (2014), the observed submm flux is linearly dependent on the dust mass as long as the emission remains optically thin. In this regime
the dust mass can be expressed as:
\begin{eqnarray}\label{eq:tommy}
M_{\mathrm{dust}} & = & \frac{S_{\nu_{obs}}}{\kappa_{\nu_{r}}\,B_{\nu_{r}}(T_{d})}\,\frac{D_{L}^{2}}{(1+z)},\  \mathrm{or} \nonumber
\ \\
\frac{M_{\mathrm{dust}}}{M_{\odot}} & = & 5.7\times10^{7}\,\left(\frac{S_{\nu_{obs}}}{\mathrm{mJy}}\right)\,\left(\frac{\nu_{obs}}{\mathrm{350\,GHz}}\right)^{-(\beta+3)}\,\left(1+z\right)^{-(\beta+4)}\left(\frac{D_{L}}{\mathrm{Mpc}}\right)^{2} \nonumber \\ 
& & \times\ \left(\exp({h\nu_{r}/kT_{d}}) - 1\right),
\end{eqnarray}
where $D_{L}$ is the luminosity distance, $\nu_{r}=\nu_{obs}(1+z)$ is the restframe frequency and $\kappa_{\nu_{r}}$ is the dust mass
absorption coefficient. The dependance on the dust temperature comes through the $\exp(h\nu_{r}/kT_{d})$ term, leading to a comparatively
modest effect on M$_{\mathrm{dust}}$ from the assumed dust temperature. For instance, a change from $T_{d}=30$\,K to 45\,K results in a
dust mass lower by a factor $\sim$2. Another uncertainty is associated with the notorious difficulty of measuring the dust mass absorption coefficient
$\kappa_{\nu}$. Estimates range from $0.07-0.15$ m$^{2}$\,kg$^{-1}$ (Hildebrand 1983; James et al. 2002). The frequency dependence has been
found to be in the range $\beta=1.5-2$ (e.g. Hughes et al. 1997; Wiklind \& Alloin 2002; Clements et al.  2010). A recent result using Planck data
found $\beta=1.78$ for Galactic dust (Planck collaboration 2011).
Here we use $\beta=1.8$, giving $\kappa_{\nu_{r}}\approx0.13\,(\nu_{obs}(1+z)/350\,\mathrm{GHz})^{1.8}$\ m$^{2}$\,kg$^{-1}$ .
The $\kappa_{\nu}$ parameter introduces an uncertainty of a factor  $\sim$2-3 in the dust mass estimate.

The dust temperature of SMGs increases with the FIR luminosity, with typical values $T_{d}\approx35\pm5$\,K for the more luminous sources
(e.g. Elbaz et al. 2011; Magnelli et al. 2012; Wardlow et al. 2011; Swinbank et al. 2013). Using the observed 870$\mu$m fluxes as listed in
Table~\ref{tab:sources}, and assuming a dust temperature of $T_{d}={\bf35}$\,K, we derive the dust masses listed in Table~\ref{tab:sfr}.
The dispersion of the derived dust mass of the SMGs is small, with M$_{\mathrm{dust}}=(4.0\pm1.3)\times10^{8}$\,M$_{\odot}$, where the
error estimate is for the sample standard  deviation and does not include uncertainties in the parameters used in the estimation.
If we knew the dust-to-gas mass ratio for the SMGs, we could use the estimated dust masses to derive the total gas mass. This ratio is not
known for SMGs but Draine et al. (2007) derived an average dust-to-gas mass ratio of 0.0052 for local galaxies with metallicities close to Solar,
showing a small dispersion in the values. Given that the SMGs in our sample appear to have close to Solar metallicities, we will
assume that this gas-to-dust mass ratio is applicable to our SMG sample as well and can be used to obtain an estimate of the total gas mass
(see also Fig. 4 in James et al. 2002).

The molecular gas masses can now be estimated if the $M_{\mathrm{H_2}}/M_{\mathrm{HI}}$ ratio is known, which is not the case for the SMG
population. Scoville (2012; 2014) compiled a sample of star forming galaxies, including ULIRGs, with submm data as well as with known molecular
and atomic gas masses. The average H$_2$/HI ratio for this sample is $\sim$2 and we will assume that this is the case for our SMGs as well.
The resulting molecular gas masses are listed in Table~\ref{tab:sfr}, where the average H$_2$ mass is
$M_{\mathrm{H_2}}=(5.1\pm1.7)\times10^{10}$\,M$_{\odot}$. Despite the assumptions needed to derive these values, the average value
is close to the average $\sim3\times10^{10}$\,M$_{\odot}$ found from direct CO observations of SMGs (e.g. Greve et al. 2005; Tacconi et al.
2008; Coppin et al. 2009; Bothwell et al. 2013 and references therein)). The only SMG in our sample with CO data (LESS 73) has an estimated 
molecular gas mass of $1.6\times10^{10}$\,M$_{\odot}$ (Coppin et al. 2010). Estimating the molecular gas mass using the 870$\mu$m flux
and the dust-to-gas mass ratio, we estimate a molecular gas mass of $5.2\times10^{10}$\,M$_{\odot}$ for LESS 73.

The molecular gas fraction, defined as $f_{H_2}=M_{H_2}/(M_{gas}+M_{*})$, is not exceptionally high, except for the two low-mass SMGs, LESS 10
and LESS 34. While these two have a molecular gas fraction of $\sim$0.65, the average gas fraction for the remaining 8 SMGs is $0.28\pm0.07$
(sample standard deviation). With our assumption of the M$_{H_2}$/M$_{\mathrm{HI}}$ ratio (see above), the total gas fraction becomes
$f_{\mathrm{gas}}=1.5\,f(H_2)$. This means that $f_{\mathrm{gas}}$ is close to unity for LESS 10 and LESS 34, while the remaining SMGs have
$f_{\mathrm{gas}}\approx0.4$. The high gas fraction found in LESS 10 and LESS 34 sets them apart form the other SMGs and could indicate that
these galaxies are in an earlier stage of the SMG phase, or that the assigned counterparts are not the real ones (see Sect.~\ref{nature} for a
discussion on the nature of these low-mass SMG counterparts).
Tacconi et al. (2013) recently presented CO observations of 52 massive main sequence galaxies at $z\sim$1.2 and $z\sim$2.2. They defined
the gas fraction as $f^{\prime}_{H_2}=M_{H_2}/(M_{H_2}+M_{*})$ and found that $f^{\prime}_{H_2}\approx$0.33 at $z\sim$1.2 and 0.47 at $z\sim$2.2.
Using this definition of the gas fraction for the LESS sources, we find that LESS 10 and LESS 34 have a gas fraction close to 100\%. The average of
the remaining 8 LESS sources, however, is $f^{\prime}_{H_2}=0.3\pm0.1$, consistent with the values found by Tacconi et al. (2013).

An alternative approach to measuring the gas mass, which by-passes the uncertainty associated with the dust mass absorption
coefficient $\kappa_{\nu}$ and the gas, is to empirically derive the proportionality between the 870$\mu$m specific luminosity and the total gas mass.
This approach has been developed by Scoville et al. (2014) using a sample of local star forming galaxies, including ULIRGs.
This allows the gas mass of high redshift star forming galaxies to be estimated from the submm flux to within a factor of a few. Using the approach
of Scoville (2014) with the empirically derived dust opacity per unit gas mass leads to gas and dust masses that are a factor 1.6 higher than those
derived using Eq.~\ref{eq:tommy}.

The star formation rate (SFR) of SMGs is usually derived using the far-infrared luminosity, based on the assumption that the dust emission is
powered by young and massive stars, possibly with a contribution of an embedded AGN. The far-infrared luminosity, however, has a strong
dependance on the dust temperature (L$_{\mathrm{FIR}}\propto T_{d}^{5-6}$) and unless the far-infrared/submm SED is determined with high
precision, especially around the peak, the resulting SFR can be highly uncertain. The single 870$\mu$m flux available for each of the SMGs
in our sample is thus not sufficient for deriving the L$_{\mathrm{FIR}}$, and subsequently the SFR.
We can, however, use a correlation derived from dust radiative transfer performed on hydrodynamical simulations of both isolated and
merging galaxies that relates the submm flux, the dust mass and the SFR (Hayward et al. 2013) to obtain an approximate estimate of the star
formation rate of the SMG's. The parametrization is valid for $S_{850}<15$\,mJy and for $z\approx1-6$:
\begin{eqnarray}\label{eq:hayward}
\mathrm{SFR} & = & 9\,\left(\frac{S_{850}}{\mathrm{mJy}}\right)^{2.33}\,\left(\frac{M_{\mathrm{dust}}}{10^{9}\ \mathrm{M}_{\odot}}\right)^{-1.26},
\end{eqnarray}
from (Hayward et al. 2013). Since the submm flux $S_{850}$ is here also used in the estimates of the dust mass, this means that the SFR, as expressed here,
is approximately proportional to $S^{1.1}_{850}\times D_{L}^2$.
Using the M$_{\mathrm{dust}}$ values derived using Eq.~\ref{eq:tommy} and the ALMA submm fluxes listed in Table~\ref{tab:sources} results in
the star formation rates listed in Table~\ref{tab:sfr}. Assuming an uncertainty in the dust mass estimates of a factor 2, and the intrinsic accuracy of
Eq.~\ref{eq:hayward} of 0.15 dex (Hayward et al. 2013), the estimated SFR should be good to within a factor $\sim$3.
The SFR values range from 460\,M$_{\odot}$\,yr$^{-1}$ (LESS 79.2), to $\sim$2.5$\times$10$^{3}$\,M$_{\odot}$\,yr$^{-1}$ (LESS 13), with
an average of $(0.8\pm0.7)\times10^{3}$\,M$_{\odot}$\,yr$^{-1}$.

While the estimates of the SFR of the SMGs rests on both the dust mass and the model results of Hayward et al. (2013),
and therefore may have accumulated errors, we will use the derived SFR's and invert the empirical relation between SFR and L$_{\mathrm{FIR}}$
(e.g. Murphy et al. 2011) to derive a rough estimate of the far-infrared luminosity for the LESS sources.
With L$_{\mathrm{FIR}}$/L$_{\odot}=6.7\times10^{9}$\,(SFR/M$_{\odot}$\,yr$^{-1}$), the L$_{\mathrm{FIR}}$ values range from
$(0.3-1.7)\times10^{13}$\,L$_{\odot}$, with an average $(1.0\pm0.4)\times10^{13}$\,L$_{\odot}$ (Table~\ref{tab:sfr}).  
This suggests that the SMGs on average belong to the class of hyperluminous infrared galaxies (HyLIRGs) and
are $\sim$5 times more luminous than local ULIRGs. The only SMG in our sample that clearly qualify as a ULIRG in terms of its far-infrared luminosity
is the merger LESS 79.2, with L$_{\mathrm{FIR}}=3\times10^{12}$\,L$_{\odot}$.
It is interesting to note that our values of the average SFR and L$_{\mathrm{FIR}}$ are $\sim$2-3 times higher than those inferred from SMGs
in the CDFS field using 3-component dust SED fitting of ALMA and Herschel data (Swinbank et al. 2013). The average SFR and L$_{\mathrm{FIR}}$ derived by
Swinbank et al. (2013) are 300\,M$_{\odot}$\,yr$^{-1}$ and $3.0\times10^{12}$\,L$_{\odot}$, respectively. These values are similar to low-z
ULIRGs (cf. Sanders \& Mirabel 1996). 

In Fig.~\ref{fig:ssfr} we compare the specific star formation rates ($SSFR=SFR/M_{*}$) of the SMGs with those of galaxies from the CANDELS UDS and GOODS-S
fields. The stellar mass of the CANDELS galaxies has been derived in the same manner as for the SMGs, but the SFRs are derived
using the dust corrected restframe 1500{\AA} luminosity. While the SSFR values for the CANDELS galaxies show a large scatter for
all stellar masses, the median values in mass bins of 0.2 dex are remarkably constant from $\log(M_{*}/M_{\odot})=8$ through
$\log(M_{*}/M_{\odot})=10$. For higher masses, the median values decline rapidly. The SFR's of the SMGs are typically 2 orders of
magnitude larger than the median galaxy for the corresponding mass bin. In the case of the two low-mass counterparts, LESS 10 and
LESS 34, the difference approaches 3 orders of magnitude in SFR. The location of the SMGs in Fig.~\ref{fig:ssfr} will not change substantially
even if the SFR values are off by a factor $\sim$3.

\section{Discussion}\label{discussion}

\subsection{The SMG counterparts}\label{identification}
Submm interferometry represents the least biased method at our disposal to identify the counterparts of SMGs. In contrast to radio
continuum as well as near- and mid-infrared emission techniques, submm interferometry targets the same emission region as initially
detected using single-dish bolometer arrays. The interferometric method, however, relies on bolometer surveys to select the targets
in the first place, and as such is not completely free of selection biases. Bolometer surveys result in flux limited samples at a wavelength
of $\sim$870$\mu$m, and thus consist of the most luminous far-infrared sources, with dust temperatures that allows the detection of the
Rayleigh-Jeans part of the continuum. Galaxies dominated by a hot dust component, for instance, may not be part of the sample
(e.g. Magnelli et al. 2010).

The identification of the SMG counterparts can be fraught with ambiguities despite using high quality submm interferometry and the
deepest and  highest angular resolution optical/near-infrared images available. In the SMG sample presented here, there are several
instances where the submm emission region is clearly separated from the optical/near-infrared emission. In some cases, this simply
reflects a real offset between a pre-existing stellar component and the region that is currently forming stars. Such offsets are present in
some local interacting/merging galaxies, such as the Antennae Galaxy, but typically not for local ULIRGs, where optical/near-infrared,
dust and molecular line emission originates in a very compact region in the center of the ULIRG (e.g. Sanders \& Mirabel 1996).

Despite the ambiguities in defining the actual counterpart to SMGs, the use of  interferometers to directly associate the submm emission
with optical/near-infrared counterpart(s) and/or nearby galaxies gives us a higher confidence in the redshift determination, regardless
whether it is spectroscopic or photometric redshifts. At least two of the SMGs in the sample presented here are located at z$>$4,
with one spectroscopically confirmed source. Given the small sample, we regard this as consistent with the high $z>4$ SMG fraction
suggested in earlier surveys (e.g. Aravena et al. 2010; Smolcic et al. 2012a,b; Walter et al. 2012; Bothwell et al. 2013; Vieira et al. 2013).
The reminder of our SMG sample falls in the `traditional' SMG redshift range $z=1.65 - 2.60$.

\subsection{Stellar properties}\label{stellar}
The SMGs in our sample appear to have two types of counterparts in terms of stellar mass. One more massive, with an average
stellar mass of $0.9 \times 10^{11}$\,M$_{\odot}$, and one type with a stellar mass 2-3 orders of magnitude lower. The distribution
of stellar masses is shown in Fig.~\ref{fig:hist} and shows the remarkably small dispersion for the high mass galaxies.
The low mass of LESS 10 and LESS 34 is not an artifact of the SED fitting process; their optical and near-infrared fluxes are 
intrinsically lower than the rest of the SMG sample while their submm properties are similar. It is tempting to view these two galaxies as
being in an earlier evolutionary stage, where the main part of the stellar mass is about to be formed (see Sect~\ref{nature}).
Our estimates of the stellar masses for the high-mass SMGs are smaller than that of the galaxies believed to be the final descendants,
giant elliptical galaxies, and are thus consistent with continued build-up of stellar mass through gas infall and/or merging. 

One perplexing characteristic of the SMG stellar population is their relatively old age (up to 1-2 Gyr). The SFR's inferred
from the far-infrared luminosity of a typical SMG can build up a $10^{11}$\,M$_{\odot}$ galaxy from scratch in $\sim10^{8}$
years, and one would naively expect the SMGs to be dominated by a young stellar population. The presence of the old(ish) stellar
population suggests that the star formation history of SMGs is more complex than envisioned. The parametrization of the star
formation history used in the SED fitting process can strongly influence the results (e.g. Michalowski et al. 2012). The `standard'
exponentially declining parametrization is a poor fit to a more complex star formation history. The delayed-$\tau$ parametrization
of the star formation history used in this paper is an improvement, allowing an active and on-going star formation period, including
an instantaneous burst. It is, however, not able to model a multiple-burst star formation history. Multiple star forming periods are
usually not used in SED fitting procedures because of the limited number of photometric data points. With too many parameters
the number of degrees of freedom approaches zero and the goodness-of-fit eventually becomes meaningless. If the real population is
composed of a young and an old stellar component, the simple, one-component star formation history will create a luminosity
weighted average stellar population. The  ages implied by the SED fit will then underestimate the age of the evolved population
while overestimating the young population. Hence, the stars in the SMG presented here could very well have a component that
is even older. This could as well have an effect on the derived stellar masses. As the stellar population ages, the M/L ratio
increases and for a fixed luminosity, the stellar mass increases. Targett et al. (2013) used a two-component stellar population
and found stellar masses that on average are $\sim$4 larger than the ones derived by us (see Table~\ref{tab:targett}). 
It is currently not clear whether this discrepancy is due to the use of one versus two stellar components, or simply reflects
differences in the SED fitting algorithms (e.g. Michalowski et al. 2012).

\begin{figure}[t]
\epsscale{0.9}
\plotone{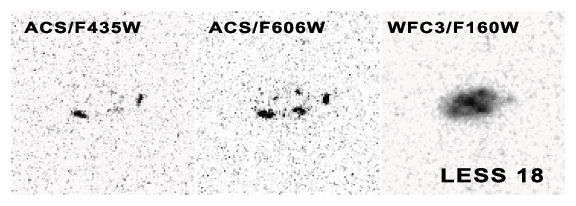}
\caption{LESS 18 shown in the ACS/F435W, F606W and WFC3/F160W filters. Each frame is about $5'' \times 5''$. The bright regions seen in
the F435W filter are either weak or absent in the F160W filter. The photometric redshift for LESS 18 is $z=1.85$ and the F435W filter thus
probes emission at $\sim1500$\AA. The lack of this emission in the F160W filter shows that these are UV bright actively star forming regions
with less dust obscuration than the rest of the galaxy. } 
\label{fig:LESS18}
\end{figure}

\begin{figure}
\epsscale{0.95}
\plotone{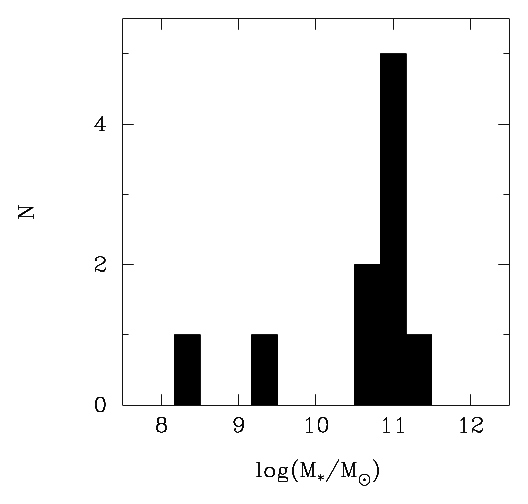}
\caption{The distribution of stellar masses of the SMGs in our sample. The two lower mass galaxies are LESS 10 and 34. The remaining 8 SMGs
show a remarkably small dispersion of their stellar masses centered on $0.9 \times 10^{11}$\,M$_{\odot}$.} 
\label{fig:hist}
\end{figure}

\begin{figure}
\epsscale{0.90}
\plotone{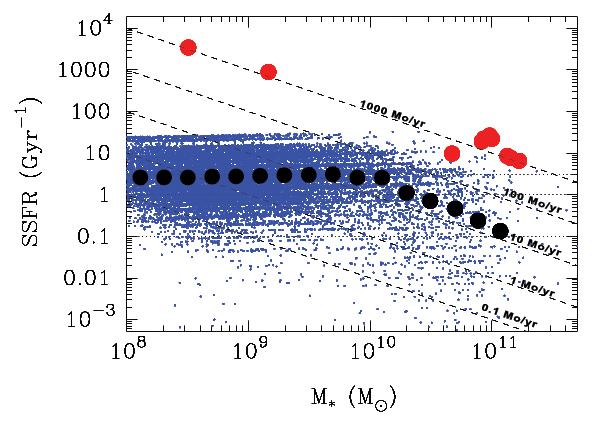}
\caption{The specific star formation rate (SSFR) as a function of stellar mass. The small blue points are galaxies from the CANDELS UDS and GOODS-S fields in the
redshift range $1.6<z<4.8$. The black circles are the median SSFR across mass bins of width 0.2 dex. The SMGs are shown as filled red circles.
The dashed lines represent the star formation rate, starting with SFR=1000 M$_{\odot}$\,yr$^{-1}$ at the top and then 100, 10, 1 and 0.1
M$_{\odot}$\,yr$^{-1}$.  The SFR for the SMG's is derived from their submm flux and estimated dust mass (see text).} 
\label{fig:ssfr}
\end{figure}

Apart from a stellar population consisting of distinct age groups, variation of other parameters across a galaxy can introduce
a bias in the parameters derived in the SED fitting process. For instance, if a galaxy consists of regions with different amounts of
 dust extinction, the SED fit will reflect the averaging over the different regions. That this is affecting at least one of the SMGs in our sample
 is illustrated in Fig.~\ref{fig:LESS18}, which shows the LESS 18 source in the ACS/F435W, F606W and the WFC3/F160W filters.
 The photometric redshift of LESS 18 is $z=1.81$ and the F435W filter probes restframe $\sim1500$\AA. The galaxy has a very
 different appearance in restframe UV compared to restframe V band (probed by the WFC3 F160W filter).
The two bright regions seen in the F435W filter are either absent or very weak in the WFC3 F160W filter showing that these are young
star forming regions with little dust obscuration, while the main component of the galaxy either contains an older stellar population
and/or is more obscured. In this case the SED parameters are biased towards a younger and less massive stellar component than
what would have been derived without the UV component. LESS 18 seems to be the only case in the present sample where there
is a UV component that is not seen at longer wavelengths.

The high metallicities of the SMGs in our sample imply a prolonged star formation activity and are in accord with the older ages of the
stars found through the SED fitting. There are, however, degeneracies between metallicity, age and dust extinction. If the ages are
overestimated, it requires either the metallicity to be even higher or the dust extinction to be more severe in order for the stellar templates
to fit the observed fluxes. Since both the implied metallicities and extinction values are already very high for the present SMG sample,
it is difficult to include a significantly younger stellar population and still achieve a good fit to the observed data. This suggests that the
young stars are likely to be heavily obscured and their contribution to the observed fluxes are small.

\subsection{Morphology}\label{morphdiscussion}
One of the main questions regarding SMGs is if they are the result of the merging of two gas-rich galaxies, with the end result being a 
giant elliptical galaxy (e.g. Blain et al. 2002), or if the star formation activity is fueled through some other mean. In the merging scenario one
would expect the SMG counterparts to show the presence of close neighbors and/or tidal features.

The SMGs presented here have some properties similar to local ULIRGs: large dust and gas-masses, a high star formation rate and a
large far-infrared luminosity (Sect.~\ref{gas}). However, only 3 of the SMGs in our sample show clear evidence for merging while 7 of them are either
completely isolated (5) or have one or two neighbors (2) within 30\,kpc but show no signs of interaction (Table~\ref{tab:results}).
This makes the SMGs different from local ULIRGs, which are almost exclusively major mergers (e.g. Sanders \& Mirabel 1996). The local
ULIRGs are powered by a combination of a circumnuclear starburst and AGN activity, fueled by a reservoir of molecular gas. In contrast,
only one of the 3 SMGs showing evidence for merging contains an AGN (LESS 13). Of the remaining 7 non-merging SMGs, 
5 show evidence of an AGN.

The fact that the majority of the SMGs appear to be isolated galaxies and only 30\% are merger (and of the three mergers, only one is
an equal mass merger), suggests that the SMGs are different from local ULIRGs, despite having similar
properties in terms of star formation rate, dust and gas content. In fact, the type of isolated massively star forming systems implied by
the results for our SMG's at high redshift are not present in the local universe.
Hydrodynamical simulations of star forming galaxies at z$\sim$2 (Dekel et al. 2009) suggests that about half of the SMGs with observed
870$\mu$m fluxes $\ga$5\,mJy are fueled by mergers with a mass ratio $>$1:10. The other half is fueled by a smoother inflow that includes
mergers with mass ratios $<$1:10. Given the small sample presented here, our results for the merger fraction and mass ratios are consistent
with the results of Dekel et al. (2009).
It is also possible that we are viewing the SMGs during different stages in their evolution, where the obvious merger systems
represent an earlier phase. However, the obvious candidates for the early part of the SMG phase are LESS 10 and LESS 34,
neither of which shows any evidence of merging. Also the gas, dust and SFR properties
(Sect.~\ref{gas}) are remarkably similar for the early candidates, the mergers and the `isolated' SMGs.
High angular resolution molecular line observations might provide an answer to these questions by mapping the distribution and
kinematical structure of the star forming gas.

Comparing the asymmetry parameter of the SMGs with galaxies of similar mass and redshift shows that the SMGs
are on average more asymmetric, but does not distinguish whether they are mergers or just exhibits a clumpy structure. The dusty
nature of SMGs could be, at least partially, responsible for the high asymmetry seen at restframe optical wavelengths. The fact that
z$\sim$2 Lyman break galaxies have properties similar to both SMGs and local ULIRGs in the M$_{20}$-Gini plot, while having very
low asymmetry values, opens up the possibility that SMGs are disk like structures, similar to the LBGs, with a high asymmetry caused
by either a clumpy structure or patchy dust distribution. 
Several studies using decomposition of the light profile of SMGs have shown them to be dominated by exponential disks (Targett et al.
2011, 2013) with an average Sersic index  $<n>\approx1.4$.  Swinbank et al. (2010) find a slightly larger Sersic index
$<n>\approx 2$ for a sample of 25 SMGs observed with HST/NICMOS. These results suggest that the light profile of SMGs in general
have a disk-like characteristic.

Kartaltepe et al. (2012) recently presented an analysis of the morphological parameters of a sample of 52 z$\sim$2 ULIRGs using
CANDELS data. The ULIRGs were selected based on their far-infrared properties as observed with Herschel. The morphology of the
sample was classified through visual inspection. The merger fraction was found to be high, $\sim$72\% compared with $\sim$32\%
for comparison sample with lower far-infrared luminosity. They also found that overall the fraction of disk-like systems is approximately equal
to the fraction of spheroids, but that the fraction of disks decreases with increasing $L_{\mathrm{FIR}}$.
Hence, the low merger fraction found for the SMGs is perplexing. Based on a comparison with the Kartelpe et al. sample, the fraction of
disk-like system should be very low for the SMGs, which is contradictory to the results found by Targett et al. (2013). High angular
resolution kinematical data are needed to determine the morphology of the SMGs.

\subsection{The nature of the SMG and their counterparts}\label{nature}
Our results regarding the properties of the SMGs and their counterparts show that in terms of their far-infrared, star formation and interstellar gas characteristics,
they constitute a homogeneous sample, but in terms of their stellar properties they divide into two distinct groups; a low stellar mass group and a high
stellar mass group. The latter has a typical stellar mass of $0.9\times10^{11}$\,M$_{\odot}$ with a remarkably small dispersion (Fig.~\ref{fig:hist}).

The possibility that the low stellar mass of LESS 10 and LESS 34 is due to $\sim$99\% of the stars being completely obscured by dust
is unlikely as their far-infrared luminosity and gas properties are similar to the other SMGs. If less UV/optical light
`leaks' out, it should be absorbed and re-radiated by dust, resulting in a higher far-infrared luminosity. The L$_{\mathrm{FIR}}$ of LESS 10
and LESS 34 are typical for SMGs, with $(7-9)\times10^{12}$\,L$_{\odot}$.
In fact, LESS 10 and LESS 34 appear to be completely `normal' SMGs in their FIR, gas and SFR properties (see Table~\ref{tab:sfr}). It is
only the stellar mass of the designated counterpart that makes them special. The SED fit of LESS 10 and LESS 34 are greatly improved by
removing a power-law continuum in the IR (Sect.~\ref{photoz}). The magnitude of the implied power-law is strong enough that essentially
all of the IRAC fluxes are removed. Performing the SED fit for LESS 10 and LESS 34 without the IRAC data, and without a continuum power-law,
also improves the $\chi^{2}$ values, while leaving the fit parameters essentially unchanged; in particular, the photometric redshift and stellar mass remain
exactly the same. This suggests that the IRAC fluxes are anomalous and inspection
of the SEDs (Fig.~\ref{fig:sed_a}) shows that the IRAC fluxes of both LESS 10 and LESS 34 appear to be disjoint from the optical/NIR fluxes.
In the case of LESS 10, there is a neighbor galaxy 0\ffas8 away, which leads to the possibility of blending of the IRAC photometry.
The SED fit of the neighbor, both with and without the IRAC fluxes, suggests that it is situated at a completely different redshift and, hence, not physically
related to the designated counterpart of LESS 10. Nevertheless, we did a SED fit combining the fluxes for the two galaxies. The result is a very poor SED
fit ($\chi^{2}_{\nu}\sim$20), with a stellar mass $\sim$5$\times10^{9}$\,M$_{\odot}$, still $\sim$20 times less massive than the high-mass SMG
counterparts. In the case of LESS 34 there is no nearby galaxy that can cause blending of the IRAC fluxes. 

The anomalous optical/NIR vs. IRAC fluxes of LESS 10 and LESS 34 could be explained if the submm and IRAC fluxes are associated with
a background galaxy, not visible in any of the optical/NIR bands. While LESS 34 has a small offset between the optical/NIR and submm
emission (although only 0\ffas34), the submm emission of LESS 10 is very well centered on the designated counterpart (Fig.~\ref{fig:smooth}).
The only way of resolving this issue is to obtain spectroscopic data on molecular or atomic fine-structure lines.

Another possibility that has to be considered is that these two galaxies represents an early stage in the development of SMGs. The characteristic age of LESS 34
is the youngest of our sample (50\,Myr), but LESS 10 is better fit by an older stellar population. The stellar age is, however, not a robust
parameter and although it can be used as a diagnostic for a reasonably large sample, it should not be used for the analysis of individual galaxies.
If all of the molecular gas in LESS 10 and LESS 34 is turned into stars, it would result in a stellar mass of $5\times10^{10}$\,M$_{\odot}$ in $\sim$40\,Myr,
the gas consumption time scale. If the star formation efficiency (SFE) is less than 100\%, the resulting stellar mass would be lower by a factor of a few.
Even with a high SFE, the resulting SMG counterpart would be at least 2-3 times less massive than the typical stellar mass of the other SMG counterparts. 
Furthermore, the high-mass SMGs are characterized by stars significantly older than 40\,Myr.
In addition, the high mass SMGs contain as much molecular gas and have as high L$_{\mathrm{FIR}}$ as LESS 10 and LESS 34 and, hence
do not seem to simply represent a later stage in the SMG episode.
In order for LESS 10 and LESS 34 to transform into the high mass counterparts, they would have to undergo at least 2, possibly 3-4,
episodes of similar star formation activity as they are currently experiencing. If these episodes occur with a time lag of $\sim$0.5\,Gyr, the
resulting galaxy would resemble the high mass galaxies. While the current data and results do not prove such episodic SMG activity, it is
consistent with such a scenario. If correct, a larger sample of SMGs would yield counterparts with stellar masses
ranging from the $10^{8}$\,M$_{\odot}$, all the way up to $10^{11}$\,M$_{\odot}$, and the apparent dichotomy of stellar masses found
here will disappear.

The presence of  low-mass counterparts to SMGs has not been seen in other studies of SMG stellar masses (cf. Wardlow et al. 2011;
Michalowski et al. 2012; Simpson et al. 2013). Hainline et al. (2011), however, did find a few cases where the stellar mass of the counterparts
approached what we find for LESS 10. Some of these studies used a simplified estimate of the stellar mass, based only on the H-band luminosity
and a M/L ratio from stellar evolution models, and the identification of the counterparts was in some cases based on data with less angular
resolution than used here.
The combination of ALMA and HST data makes the
identification of the SMG counterpart more precise than previously, and our stellar mass estimates have proven to be a robust outcome from the
SED fitting.

The above discussion begets the question of what constitutes a `SMG episode'? Merger between gas-rich galaxies or, a more or less continuous
accretion? The general assumption is that the most luminous SMGs are powered through a merger while less luminous SMGs are fueled through
a more continuous accretion, possibly involving accretion of clumps and dwarf galaxies with a mass ratio $<$0.1 (cf. Finlator et al. 2006; Dekel
et al. 2009; Dav\'{e} et al. 2010). This is supported by the data presented here.
The only verified equal mass merger in our sample is LESS 13, which is also the most far-infrared luminous SMG. The two other mergers (LESS 67
and LESS 79.2) have mass ratios of 1/5 and 1/20, respectively, and both have far-infrared luminosities below the average of our sample. In fact,
LESS 79.2 is the least luminous object of our SMGs. If this type of un-equal mass mergers/accretion is fueling the star formation activity on time scales
less than $\sim$40\,Myr, the gas is constantly replenished and stellar mass is built-up in a more or less continuous fashion.
The gas accretion rate at z$\ga$2 is estimated to be sufficient to sustain a SFR of several $10^{2}$\,M$_{\odot}$\,yr$^{-1}$ (e.g. Dekel et al.
2009, 2013).
At some point, however, the star formation is quenched and this seems to occur around $(1-2)\times10^{11}$\,M$_{\odot}$ as more
massive SMG counterparts are not found, or at least appear to be very rare.

\section{Summary}

Using sensitive and high angular resolution HST data from the CANDELS and GOODS  projects we identify optical counterparts
of 10 sub-millimeter sources recently observed with ALMA and contained within the CANDELS coverage of the GOODS-S field.
The sources are all submillimeter galaxies with extraordinarily high star formation rates and contain large amounts of dust.
Only one SMG, LESS 79.4, remains unidentified in the HST data. This source has a very faint submm continuum and may be
a spurious submm source. Two additional SMG sources listed in Hodge et al. (2013), LESS 17.2 and LESS 17.3, fall outside
the ALMA primary beam and neither is associated with an optical/NIR counterpart and could therefore also be surious submm sources.

Our sample of 10 SMGs contains 3 mergers, where one is an equal mass merger. The remaining 7 SMGs are isolated, with 2
having neighbors within 30 kpc but with no evidence of gravitational interaction. While the SMGs share similarities with local
ULIRGs in terms of gas and dust content as well as star formation rates, they differ from the local ULIRGs by being mostly isolated
systems. This type of isolated far-infrared luminous, dusty, gas rich and massively star forming galaxies seen at high redshift is not
present in the local universe, but is consistent with models showing that high-z SMGs are fueled by both smooth inflows and major
mergers, depending on the observed intensity if the star formation activity.

Comparing the morphological parameters of the SMGs with a sample of galaxies at the same redshift and with similar stellar masses
shows that the SMGs are among the most asymmetric galaxies in the sample. Apart from the merging systems, the asymmetry can be caused
by a clumpy light distribution, clearly seen in three of the isolated galaxies. Asymmetry can also be caused by a patchy dust distribution,
although this is not directly inferred from the present data.

In most cases the center of the submm emission, as derived from the ALMA data, is directly associated with the restframe optical
emission, taking the positional uncertainties into consideration, but for a few of the SMGs the submm emission is offset from the
stellar component by $\sim$4 kpc. The SMGs with larger offsets include 2 of the merger systems and one isolated SMG. In the
latter case, it cannot be excluded that the submm emission originates from  a background galaxy. In general, despite the high
angular resolution provided by HST and ALMA, the definition of the optical counterpart can remain elusive and great care has
to be applied when comparing properties such as star formation rate, stellar mass and age of the SMG and the optical counterpart.

SED fitting to the SMGs provide their photometric redshift, stellar mass, extinction, metallicity and star formation history. The redshifts of
the SMGs range from z$=1.65$ to z$=4.76$. The sample contains four SMGs with spectroscopic redshifts (including the highest redshift
source at z$=4.76$). Two out of ten SMGs are located at z$>$4. The median stellar mass is $9 \times 10^{10}$\,M$_{\odot}$, lower than
early estimates of the characteristic stellar mass of SMGs. The age of the stellar populations is typically $\sim$0.7$-$1 Gyr. This is well
in excess of the expected length of the SMG phase and suggests a complex star formation history involving previous star formation epochs.
The metallicity as well as the dust extinction is already large for the SMGs in the sample, making it difficult to explain the old ages with
degeneracies among the fitted parameters.

The SMGs can be divided into two groups in terms of their stellar masses. One high-mass group with stellar masses
$\sim 1 \times 10^{11}$\,M$_{\odot}$ showing a remarkably small dispersion, and one low-mass group with stellar masses
2-3 orders of magnitude smaller than the high-mass group. Both groups have similar far-infrared/submm properties. It is possible
that the low-mass SMG counterparts represent an earlier stage of the SMG phase and stellar mass will rapidly increase with time.
The two galaxies with the smallest stellar masses also have the largest contribution from a mid-IR power-law
continuum. The small stellar mass is, however, not caused by the removal of a mid-IR power-law as the optical through near-infrared
fluxes are lower than for the high-mass SMG counterparts. The low mass counterparts will not be able to become the
high mass SMG counterparts without additional infall of gas, and with episodic star formation spread out over several $10^{8}$ years.
We also consider the possibility that the submm emission is not associated with the designated counterparts, but originates in
a background source, which remains completely  invisible at optical/NIR wavelengths.

Star formation rates as well as dust and gas masses are inferred from the submm emission. It is found that the SMGs constitutes a
homogenous group in terms of their gas and dust properties, without showing the dichotomy seen in the stellar masses. Except for the
two low mass galaxies LESS 10 and LESS 34, for which the baryonic mass appears to be dominated by molecular gas, the 
molecular gas fraction is found to be $\sim$28\%. If the star formation rates, derived from model results, are correct,  the typical
gas consumption time scale for the SMGs is short, only $\sim37$\,Myr, and unless the gas is replenished through accretion, the SMGs
can add $\sim4\times10^{10}$\,M$_{\odot}$ of stars during the SMG phase, leading to a final stellar mass of
$\sim1-2 \times 10^{11}$\,M$_{\odot}$ for the SMG descendants.

\acknowledgements
This work is based on observations taken by the CANDELS Multi-Cycle Treasury Program with the NASA/ESA HST,
which is operated by the Association of Universities for Research in Astronomy, Inc., under NASA contract NAS5-26555.
This work is based in part on observations with the Spitzer Space Telescope, which is operated by the Jet Propulsion
Laboratory, California Institute of Technology under a contract with NASA, and also on the Very Large Telescope, operated
by the European Southern Observatory. 

This paper makes use of the following ALMA data: ADS/JAO.ALMA\#2011.0.00294.S. ALMA is a partnership of ESO (representing
its member states), NSF (USA) and NINS (Japan), together with NRC (Canada) and NSC and ASIAA (Taiwan), in cooperation with
the Republic of Chile. The Joint ALMA Observatory is operated by ESO, AUI/NRAO and NAOJ.

\clearpage

\begin{deluxetable}{lllccccccrl}
\tablewidth{0pt}
\tabletypesize{\scriptsize}
\rotate

\tablecaption{LESS sources within the GOODS-S field\label{tab:sources}}

\tablehead{
\multicolumn{1}{c}{(1)}      &
\multicolumn{1}{c}{(2)}      &
\multicolumn{1}{c}{(3)}      &
\multicolumn{1}{c}{(4)}      &
\multicolumn{1}{c}{(5)}      &
\multicolumn{1}{c}{(6)}      &
\multicolumn{1}{c}{(7)}      &
\multicolumn{1}{c}{(8)}      &
\multicolumn{1}{c}{(9)}      &
\multicolumn{1}{c}{(10)}      &
\multicolumn{1}{c}{(11)}    \\
\ \\
\multicolumn{1}{c}{IAU name}        &
\multicolumn{1}{c}{LESS ID}           &
\multicolumn{1}{c}{GOODS-S ID}  &
\multicolumn{1}{c}{RA}                    &
\multicolumn{1}{c}{DEC}                &
\multicolumn{1}{c}{$\Delta$r}         &
\multicolumn{1}{c}{z$_{phot}$}                &
\multicolumn{1}{c}{z$_{spec}$}               &
\multicolumn{1}{c}{S$_{870}$}      &
\multicolumn{1}{c}{S$_{24}$}        &
\multicolumn{1}{c}{Ref for $z_{spec}$} \\
\multicolumn{3}{c}{}                             &
\multicolumn{2}{c}{(J2000.0)}            &
\multicolumn{1}{c}{arcsec}                 &
\multicolumn{1}{c}{}                             &
\multicolumn{1}{c}{}                             &
\multicolumn{1}{c}{mJy}                      &
\multicolumn{1}{c}{$\mu$Jy}              &
\multicolumn{1}{c}{}  \\
}
\startdata
LESS J033219.0$-$275219   & LESS 10\tablenotemark{a)} &   4414   & 53.079417 & $-$27.870778 &  0.32 & 1.82 & --- & 5.2$\pm$0.5 & 148.0$\pm$4.1 & \\
LESS J033249.2$-$274246   & LESS 13      &  23751& 53.204125 & $-$27.714389 & 0.63 &  4.31 & --- & 8.0$\pm$0.6 & 19.2$\pm$5.8 & \\
LESS J033207.6$-$275123   & LESS 17.1\tablenotemark{b)}   &    5953  & 53.030417 & $-$27.855778 & 0.59 & 1.60 & 2.0353: & 8.4$\pm$0.5 & 220.0$\pm$5.5 & Balestra et al. (2010) \\
LESS J033207.6$-$275123   & LESS 17.2   &       ---    & 53.034417 & $-$27.855472 &          &          &                & 3.7$\pm$0.9 &                              & \\
LESS J033207.6$-$275123   & LESS 17.3   &       ---    & 53.030708 & $-$27.859417 &          &          &                & 5.1$\pm$1.2 &                              & \\
LESS J033205.1$-$274652   & LESS 18      &   15261 & 53.020333 & $-$27.779917 & 0.34 & 1.83 & --- & 4.4$\pm$0.5 & 620.0$\pm$6.5& \\
LESS J033217.6$-$275230   & LESS 34\tablenotemark{c)}      &   3818   & 53.074833 & $-$27.875917 & 0.35 & 1.81 & --- & 4.5$\pm$0.6 & 11.3$\pm$2.9 & \\ 
LESS J033225.7$-$275228   & LESS 45      &     3943 & 53.105250 & $-$27.875139 & 0.02 & 2.32 & --- & 6.0$\pm$0.5 & 141.0$\pm$4.0 & \\
LESS J033243.3$-$275517   & LESS 67.1   &   899    & 53.180000 & $-$27.920639 & 0.31&  2.13 & 2.122 & 4.5$\pm$0.5 & 554.0$\pm$4.4 & Kriek et al. (2008)\\ 
LESS J033243.3$-$275517   & LESS 67.2   &       ---   & 53.179250 & $-$27.920750 &         &           &            & 1.7$\pm$0.4 &                              &        \\
LESS J033229.3$-$275619   & LESS 73      &     273   & 53.122042 & $-$27.938806 & 0.23 & 4.40 & 4.762 & 6.1$\pm$0.5 & 31.6$\pm$5.1 & Vanzella et al. (2008)\\
LESS J033221.3$-$275623   & LESS 79.1  &    216    & 53.088083 & $-$27.940833 & 0.34 & 2.62 & --- & 4.1$\pm$0.4 & 42.4$\pm$5.9 & \\
LESS J033221.3$-$275623   & LESS 79.2\tablenotemark{d)}  &    282    & 53.090000 & $-$27.940000 & 0.46 & 1.66 & 2.073: & 2.0$\pm$0.4 & 586.0$\pm$6.2 & Balestra et al (2010) \\
LESS J033221.3$-$275623   & LESS 79.4  &      ---     & 53.088250 & $-$27.941806  &          &          &             & 1.8$\pm$0.5 &                              & \\
\enddata
\tablenotetext{(1)}{The IAU designation for the sources.}
\tablenotetext{(2)}{This shortened name (same as in Hodge et al. 2013) will be used to identify the sources in the paper.}
\tablenotetext{(3)}{The CANDELS catalog ID (Guo et al. 2013) for the counterparts.}
\tablenotetext{(4-5)}{Center coordinates for the LESS sources as derived from the ALMA data (Hodge et al. 2013).}
\tablenotetext{(6)}{Separation between the LESS source and the corresponding F160W object.}
\tablenotetext{(8)}{Colon marks low quality spectroscopic redshift.}
\tablenotetext{(9)}{Submm continuum flux from Hodge et al. (2013). The fluxes have been corrected for the ALMA primary beam response.}
\tablenotetext{(10)}{Spitzer MIPS 24$\mu$m data.}
\tablenotetext{a)}{Targett et al. (2013) identified the SMG counterpart with a galaxy 0\ffas8 north of the position of the submm emission.}
\tablenotetext{b)}{We include all three LESS 17.1, 17.2 and 17.3 (see Hodge et al. 2013) fro completeness. In the rest of the paper we refer the LESS 17.1 as LESS 17.}
\tablenotetext{c)}{This source was identified with a z$=$1.08 spiral galaxy in Targett et al. (2013).
The ALMA data shows it to be associated with a faint galaxy at z$_{phot}\approx$1.8.}
\tablenotetext{d)}{This source was identified as the submm counterpart for LESS 79 in Targett et al. (2013).}
\end{deluxetable}

\begin{deluxetable}{lccccccccccccl}
\tablewidth{0pt}
\tabletypesize{\scriptsize}
\rotate

\tablecaption{Parameters for the SMG counterparts derived from SED fitting\label{tab:results}}

\tablehead{
\multicolumn{1}{c}{(1)}      &
\multicolumn{1}{c}{(2)}      &
\multicolumn{1}{c}{(3)}      &
\multicolumn{1}{c}{(4)}      &
\multicolumn{1}{c}{(5)}      &
\multicolumn{1}{c}{(6)}      &
\multicolumn{1}{c}{(7)}      &
\multicolumn{1}{c}{(8)}      &
\multicolumn{1}{c}{(9)}      &
\multicolumn{1}{c}{(10)}    &
\multicolumn{1}{c}{(11)}    &
\multicolumn{1}{c}{(12)}    &
\multicolumn{1}{c}{(13)}    &
\multicolumn{1}{c}{(14)}     \\
\ \\
\multicolumn{1}{c}{LESS}               &
\multicolumn{1}{c}{GOODS-S}       &
\multicolumn{1}{c}{z} &
\multicolumn{1}{c}{E$_{B-V}$}       &
\multicolumn{1}{c}{Age}                   &
\multicolumn{1}{c}{$\tau$}               &
\multicolumn{1}{c}{Z}                        &
\multicolumn{1}{c}{$\log(L_{bol})$}  &
\multicolumn{1}{c}{$\log(M_{*})$}     &
\multicolumn{1}{c}{$\delta f_{8\mu m}$} &
\multicolumn{1}{c}{$\alpha$}  &
\multicolumn{1}{c}{$\chi^{2}_{\nu}$}  &
\multicolumn{1}{c}{AGN}  &
\multicolumn{1}{c}{Environment} \\
\multicolumn{2}{c}{}                              &
\multicolumn{2}{c}{}                              &
\multicolumn{1}{c}{Gyr}                        &
\multicolumn{1}{c}{Gyr}                        &
\multicolumn{1}{c}{$Z_{\odot}$}         &
\multicolumn{1}{c}{L$_{\odot}$}         &
\multicolumn{1}{c}{M$_{\odot}$}        &
\multicolumn{5}{c}{} \\
}
\startdata
LESS 10      &   4414   & 1.85      & 0.250 & 0.500 & 0.80 & 1.0 & 10.987 &  9.392 & 0.9 & 2.0 & 1.7 & IR & Isolated\\
LESS 13A   &  23751 & 4.20      & 0.100 & 0.800 & 0.30 & 1.0 & 11.634 &  10.981 & 0.0 & --- & 0.2 & IR & Merger (A) \\
LESS 13B   & 23757  & 4.20      & 0.025 & 1.600 & 0.60 & 1.0 & 11.418 &  10.771 & 0.2   & 2.0 & 2.7 & --- & Merger (B) \\
LESS 17      &    5953  & 2.05    & 0.450 & 0.300 & 0.20 & 0.2 & 12.656 & 11.000 & 0.0 & --- & 0.9 & X,R & Isolated \\
LESS 18      &   15261 & 1.81     & 0.350 & 3.000 &  0.10 & 2.5 & 11.461 & 11.220 & 0.2 & 2.0 & 0.9 & --- & Isolated (clumpy) \\
LESS 34      &    3818  & 1.81     & 0.425 & 0.050 & 0.40 & 1.0 & 10.877 &  8.361 & 0.8 & 2.0 & 1.3 &  --- & Neighbor \\
LESS 45      &    3943 & 2.30      & 0.525 & 0.800 & 0.40 & 1.0 & 11.932 & 10.896 & 0.2 &1.9 & 0.8 & X,IR & Neighbor (clumpy)\\
LESS 67A   &   899    & 2.12    & 0.375 & 2.000 & 0.20 & 2.5 & 12.283 & 11.125 & 0.0 & --- & 0.7 & --- & Merger (A) \\
LESS 67B   &   881    & 2.12    & 0.125 & 0.200 & 0.01 & 0.2 & 10.982 &  9.824  & 0.0   & --- & 2.9 & ---  & Merger (B)\\
LESS 73      &     273  & 4.76    & 0.250 & 0.800 & 0.80 & 1.0 & 12.265 & 10.933 & 0.0 & --- & 4.7 & X,IR & Isolated \\
LESS 79.1   &    216  & 2.60       & 0.650 & 0.500 & 0.20 & 1.0 & 11.815 & 11.158 & 0.0 & --- & 0.3 & IR & Isolated (clumpy) \\
LESS 79.2A &   282  & 1.65        & 0.425 & 0.100 & 0.30 & 2.5 & 11.958 & 10.672 & 0.0 & --- & 0.7 & --- & Merger (A) \\
LESS 79.2B &   245  & 1.65        & 0.350 & 0.100 & 0.30 & 2.5 & 11.265 &  9.979 & 0.0 & --- & 3.7  &  ---   & Merger (B) \\
\enddata
\tablenotetext{(1)}{Mergers are designated with A and B, where A is the component closest to the submm emission.}
\tablenotetext{(2)}{Galaxy ID in the CANDELS catalog (Guo et al. 2013).}
\tablenotetext{(3)}{Final redshifted adopted from Monte Carlo simulations, spectroscopic redshift or a combination of the two.}
\tablenotetext{(10)}{Fraction of the 8$\mu$m flux subtracted as a non-stellar power-law SED.}
\tablenotetext{(11)}{The slope of the power-law continuum; $f_{\nu} \propto \nu^{-\alpha}$.}
\tablenotetext{(13)}{AGN detection: $'-'=$ no AGN signature; X$=$X-ray signature; IR$=$IR signature; R$=$radio signature.}
\tablenotetext{(14)}{Environment; Isolated = no companion within 30\,kpc; Neighbor= companion within 30 kpc,
but no signs of interaction; Merger: A=main SMG counterpart, B=secondary merger component.}
\end{deluxetable}

\begin{deluxetable}{lcccc}
\tablewidth{0pt}
\tabletypesize{\scriptsize}

\tablecaption{Sources in common with Targett et al. (2013)\tablenotemark{1}\label{tab:targett}}

\tablehead{
\multicolumn{1}{c}{LESS ID}                           &
\multicolumn{2}{c}{$z_{\mathrm{phot}}$}      &
\multicolumn{2}{c}{$\log(M_{*}/M_{\odot})$}  \\
\ \\
\multicolumn{1}{c}{}             &
\multicolumn{1}{c}{TW}       &
\multicolumn{1}{c}{TT}        &
\multicolumn{1}{c}{TW}       &
\multicolumn{1}{c}{TT}         \\
}
\startdata
  18    & 1.81 & 1.65 & 11.22 & 11.78 \\
  45    & 2.30 & 2.13 & 10.90 & 11.59 \\
  67A    & 2.12 & 2.08 & 11.12 & 11.53 \\
  73    & 4.76 & 4.76 & 10.93 & 11.18 \\
  79.2A & 1.65 & 1.73 & 10.67 & 11.82 \\
\enddata
\tablenotetext{1}{TW = this paper ; TT = Targett et al. (2013)}
\end{deluxetable}

\begin{deluxetable}{lccccccccc}
\tablewidth{0pt}
\tabletypesize{\scriptsize}

\tablecaption{Morphological parameters of LESS sources within the GOODS-S field\label{tab:morph}}

\tablehead{
\multicolumn{1}{c}{(1)}     &
\multicolumn{1}{c}{(2)}     &
\multicolumn{1}{c}{(3)}     &
\multicolumn{1}{c}{(4)}     &
\multicolumn{1}{c}{(5)}     &
\multicolumn{1}{c}{(6)}     &
\multicolumn{1}{c}{(7)}     &
\multicolumn{1}{c}{(8)}     &
\multicolumn{1}{c}{(9)}     &
\multicolumn{1}{c}{(10)}  \\
\ \\
\multicolumn{1}{c}{LESS}               &
\multicolumn{1}{c}{GOODS-S}       &
\multicolumn{1}{c}{C}                      &
\multicolumn{1}{c}{$\Delta$C}       &
\multicolumn{1}{c}{A}                      &
\multicolumn{1}{c}{$\Delta$A}       &
\multicolumn{1}{c}{S}                       &
\multicolumn{1}{c}{$\Delta$S} &
\multicolumn{1}{c}{M$_{20}$}       &
\multicolumn{1}{c}{Gini}                  \\
}
\startdata
  10    &   4414 &   2.59  &  0.25  &  0.18  &  0.02  &  0.00  &  0.00  &  $-$1.68   &  0.55   \\
  13    & 23751 &   2.81  &  0.09  &  0.09  &  0.18  &  0.54  &  0.14  &  $-$0.76  &  0.92    \\
  17    &   5953 &   3.27  &  0.18  &  0.26  &  0.01  &  0.13  &  0.02  &  $-$1.99   &  0.56   \\
  18    & 15261 &   2.09  &  0.10  &  0.18  &  0.02  &  0.10  &  0.03  &  $-$1.34  &   0.51   \\
  34    &   3818 &   3.12  &  0.22  &  0.43  &  0.09  &  0.00  &  0.00  &  $-$1.74  &  0.90    \\
  45    &   3943 &   2.34  &  0.13  &  0.31  &  0.07  &  0.22  &  0.07  &  $-$1.40   &  0.64   \\
  67    &     899 &   2.44  &  0.11  &  0.56  &  0.03  &  0.16  &  0.04  &  $-$1.25  &  0.64    \\
  73    &     273 &   2.41  &  0.30  &  0.16  &  0.01  &  0.00  & 0.00  &  $-$2.07 &  0.58    \\
  79.1 &    216  &  2.04  &  0.11  &  0.32  &  0.10 &  0.28  &  0.10  &  $-$0.78 &  0.56    \\
  79.2 &    282  &  2.78  &  0.12  &  0.60  &  0.02 &  0.14  &  0.03  &  $-$1.47  &  0.65    \\
\enddata
\tablenotetext{(3)}{Concentration index.}
\tablenotetext{(5)}{Asymmetry index, $A$; a value A$>$0.35 usually indicates a merger system.}
\tablenotetext{(7)}{Clumpiness index.}
\tablenotetext{(9)}{The M$_{20}$ index measures the second order moment of the brightest regions.}
\tablenotetext{(10)}{The Gini coefficient.}
\end{deluxetable}

\begin{deluxetable}{lcccc|cc}
\tablewidth{0pt}
\tabletypesize{\scriptsize}

\tablecaption{Gas and star formation properties of the SMGs\label{tab:sfr}}

\tablehead{
\multicolumn{1}{c}{LESS}               &
\multicolumn{1}{c}{GOODS-S}       &
\multicolumn{1}{c}{M$_{\mathrm{dust}}$\tablenotemark{a)}}    &
\multicolumn{1}{c}{M$_{\mathrm{H_{2}}}$\tablenotemark{b)}}  &
\multicolumn{1}{c}{$f_{\mathrm{gas}}$\tablenotemark{c)}}  &
\multicolumn{1}{c}{SFR\tablenotemark{d)}} &
\multicolumn{1}{c}{L$_{\mathrm{FIR}}$}          \\
\ \\
\multicolumn{2}{c}{}                                             &
\multicolumn{1}{c}{$10^{8}$\,M$_{\odot}$}     &
\multicolumn{1}{c}{$10^{10}$\,M$_{\odot}$}   &
\multicolumn{1}{c}{}                                              &
\multicolumn{1}{c}{$10^3$\,M$_{\odot}$\,yr$^{-1}$}   &
\multicolumn{1}{c}{$10^{13}$\,L$_{\odot}$}    \\
}
\startdata
  10    &   4414                           & 4.1 & 5.2 & 0.65  & 1.3\tablenotemark{e)} & 0.9\\
  13    & 23751                           & 5.4 & 6.9 & 0.35 & 2.5  & 1.7 \\
  17    &   5953                           & 6.5 & 8.3 & 0.37 & 2.2  & 1.5  \\
  18    & 15261                           & 3.4 & 4.4 & 0.19 & 1.1  & 0.7  \\
  34    &   3818                           & 3.5 & 4.5  & 0.66 & 1.1 & 0.7  \\
  45    &   3943                           & 4.6 & 5.8 & 0.35 & 1.6  & 1.1   \\
  67    &     899                           & 3.5 & 4.4 & 0.22 & 1.1  & 0.8  \\
  73\tablenotemark{f)} & 273 & 4.0 & 5.2 & 0.32 & 1.9  & 1.3  \\
  79.1 &    216                            & 3.0 & 3.9 & 0.19 & 1.1  & 0.7  \\
  79.2 &    282                            & 1.6 & 2.0 & 0.26 & 0.5  & 0.3  \\
\enddata
\tablenotetext{a)}{Dust mass derived using Eq.~\ref{eq:tommy}.}
\tablenotetext{b)}{Derived from the dust mass and assuming a dust-to-gas mass ratio 0.0052 and M$_{H_2}=2/3$\,M$_{\mathrm{gas}}$.}
\tablenotetext{c)}{Molecular gas fraction = $M_{\mathrm{H_2}}/(M_{\mathrm{gas}} + M_{*})$}
\tablenotetext{d)}{Star formation rated derived from Hayward (2013) assuming M$_{\mathrm{dust}}$ derived from Eq.~\ref{eq:tommy}.}
\tablenotetext{e)}{Values to the right of the vertical line are derived using a correlation between the submm flux, the dust mass and the SFR (see text).}
\tablenotetext{f)}{M$_{H_2}=1.6\times10^{10}$\,M$_{\odot}$ from CO(2--1) observations (Coppin et al. 2009).}
\end{deluxetable}

\end{document}